\documentclass[12pt,a4paper]{article}

\usepackage{jheppub}
\usepackage{comment}
\usepackage{stackengine}
\usepackage{mathtools}
\usepackage{tikz}
\usetikzlibrary{tikzmark,calc,,arrows,shapes,decorations.pathreplacing}

\newcommand{\cM}{\mathcal{M}}

\newcommand{\cG}{\mathcal{G}}

\newcommand{\PE}{\operatorname{PE}}

\newcommand{\bdry}{\mathrm{bdry}}
\newcommand{\raw}{\mathrm{raw}}

\newcommand{\eps}{\epsilon}

\title{
Constructing the Monopole Formula for $A$-Type Good Quivers via Quiver Yangians}

\author[a,b]{Tiantai Chen}
\affiliation[a]{Institute of Theoretical Physics, Chinese Academy of Sciences,\\
	\hspace*{0.3cm}Zhongguancun East Road 55, Beijing 100190, China}
\affiliation[b]{School of Physical Sciences, University of Chinese Academy of Sciences,\\
\hspace*{0.3cm}Yuquan Road 19, Beijing 100049, China}
\emailAdd{chentiantai@itp.ac.cn}

\abstract{Based on the conjectured (and checked for tree-type quivers) quiver Yangian/Coulomb branch algebra
correspondence, we give a quiver Yangian interpretation of the monopole formulas for 3D $\mathcal N=4$ good $A$-type quiver gauge theories, also known as the $T_\rho(SU(N))$ theories. Using the algebra action on the $\frac12$-BPS vortices, we reorganize the generators of the truncated shifted quiver Yangian into boundary-adapted generators, and obtain the classical relations in terms of $U(N)$ Casimirs on the Slodowy slice $\mathcal S^{\mathfrak{gl}_N}_\rho$, whose coordinates are given by these boundary-adapted generators.  We also show that the boundary-adapted generators and the Casimir relations have the correct fugacities as predicted by the Hall-Littlewood expression of the monopole formula for $T_\rho(SU(N))$, hence reconstructing it as the Hilbert series of the truncated shifted quiver Yangian.
 }

\begin{document}
\maketitle
\flushbottom

\section{Introduction}
\label{sec:introduction}

Supersymmetric gauge theories with eight supercharges provide a setting in
which strongly coupled quantum geometry can be studied with unusual precision.
For a 3D $\mathcal N=4$ theory, the moduli space generically splits into Higgs and Coulomb branches.
The former is protected against quantum corrections and admits a hyperk\"ahler-quotient description, whereas
the latter receives quantum corrections and is generated, in a chosen complex
structure, by vector-multiplet complex scalars and $\frac 1 2$-BPS monopole operators.  This distinction
is central to the analyses of gauge dynamics
and mirror symmetry (which interchanges the Higgs branch of one theory with the Coulomb branch of the dual theory) in 3D \cite{SeibergWitten1996,IntriligatorSeiberg1996}.  Brane constructions \cite{HananyWitten1997} lead to the geometric picture for the relationship between the Coulomb branch and the monopole moduli space and for the mirror symmetry, while the symplectic duality \cite{braden2022,BullimoreEtAl2016,Kamnitzer2022} relates the Coulomb branch and the Higgs branch of the same theory as a symplectic pair.
The study of Coulomb branches sits at a
meeting point of supersymmetric field theory, symplectic geometry, and
geometric quantization theory.

\medskip

In this paper,  we will focus on the algebraic structure on the Coulomb branch of 3D $\mathcal N=4$ theory, and are particularly interested in the Hilbert series of the algebra.
There are different complementary ways of making the (quantum) Coulomb branch algebra precise.
The physical abelianization construction describes monopole operators (dressed by vector-multiplet scalars) and their
products on a birational abelian patch, and quantizes these relations by turning on an
$\Omega$-deformation \cite{BullimoreDimofteGaiotto2017,Nekrasov_2010}.
The mathematical construction of \cite{Nakajima2016,BFN2018}
defines the Coulomb branch chiral ring by an equivariant Borel--Moore homology convolution
algebra, and quantizes the algebra via geometric quantization.  For $ADE$-type quiver gauge theories this framework
connects Coulomb branches with slices in the affine Grassmannian, while the algebras are given by corresponding truncated shifted Yangians \cite{BFN2019}.  Vortex Hilbert spaces provide a third, representation-inspired
viewpoint: $\frac{1}{2}$-BPS monopoles faithfully act on $\frac 1 2$-BPS vortices, and produce Verma-type
modules of the quantized Coulomb branch algebra
\cite{BullimoreEtAl2018,BullimoreEtAl2016}.

The \textit{monopole formula} \cite{CremonesiHananyZaffaroni2014}, serving as the Hilbert series of the Coulomb branch algebra, counts the (gauge-invariant) $\frac 1 2$-BPS monopole operators dressed by the vector-multiplet complex scalars, graded by the $R$-charge of the bare monopole and Casimir dressings of the residual gauge group.
Although the Coulomb branch algebra itself is usually cumbersome to fully describe, the monopole formula is remarkably effective, extracting the graded spectrum of Coulomb branch operators, often
with full topological refinement. However, its algebraic interpretation is less
direct.  A Hilbert series, or equivalently its plethystic
logarithm, records the degrees of
generators and relations (and syzygies, if there are any) \cite{BenvenutiEtAl2006}, but it does not by itself exhibit
the corresponding operators or construct the relations.

For quiver gauge theories of $A$-type, this problem is especially sharp. 
A closed-form 
expression for the monopole formula of such
theories was given using Hall-Littlewood polynomials in \cite{CremonesiEtAl2014}, together with the
associated complete-intersection structure.  The plethystic
logarithm is hence finite and splits into positive terms and negative terms, corresponding to generators and relations respectively.
The closed-form formula therefore tells us exactly what must be
explained, but leaves open a constructive operator interpretation of its
individual terms.

\medskip

The quiver Yangian offers such a constructive language.  Quiver Yangians are generalizations of original Yangians of Lie algebra $\mathfrak g$. They  originated in \cite{LiYamazaki2020} and were first used to study the BPS algebras for non-compact toric Calabi-Yau threefolds.
Their shifted versions and realizations via supersymmetric quantum mechanics are developed in \cite{GalakhovLiYamazaki2021,GalakhovYamazaki2022}.
Extension to general quivers, and algorithmic constructions of quiver Yangian representations are discussed in
\cite{Li2024,GalakhovEtAl2024}.
For comparisons with $R$-matrix constructions and $W$-algebras see \cite{Bao2022,Bao2023}. These constructions are related to the wider
cohomological Hall algebra (CoHA) picture of BPS algebras
\cite{KontsevichSoibelman2011}.

It was recently conjectured that the quiver Yangian can also serve as a reformulation of the Coulomb branch algebra of 3D $\mathcal N=4$ theory \cite{ChenLi2026}, and this conjecture has been explicitly checked for tree-type quivers. In those cases, the quiver Yangian generators are  gauge-invariant combinations of dressed monopoles and vector-multiplet scalars, and hence provide a natural basis for constructing the monopole formula.
It therefore gives a
starting point for asking a more refined question: can one read the monopole
formula itself from the quiver Yangian, and can one interpret the
algebraic meaning of each of its terms?

\medskip

In this paper we answer this question for $A$-type quivers at a
precisely specified level.  The 3D $\mathcal N=4$ theories of interest are specified by linear quivers of \textit{good} type (in the sense of \cite{GaiottoWitten2009})
\begin{equation}
 [U(N)]-(U(N_1))_1-\cdots-(U(N_d))_d,
 \qquad N_0=N,\quad N_{d+1}=0,
 \label{eq:intro-quiver}
\end{equation}
where each round node $(U(N_a))$ corresponds to a gauge factor $U(N_a)$ together with its 3D $ \mathcal N=4$ vector-multiplet, while the square node $[U(N)]$ corresponds to the flavor group $U(N)$. An edge between two nodes represents a 3D $ \mathcal N=4$ hypermultiplet, living in the bi-fundamental representation of the group factors on its two sides, see \cite{ChenLi2026}. For later convenience, we have also labeled the gauge nodes by $a=1,\dots,d$.  The tree structure of $A$-type quivers can be seen as follows: with node 1 regarded as the root node, the tree grows in the direction of increasing $a$.

The theory is of good type, meaning that the rank differences
$L_a=N_{a-1}-N_a$ constitute a partition of $N$:
\begin{equation}
     \qquad L_1\geq L_2\geq\cdots\geq L_{d+1}>0\,,  \qquad \sum_{a=1}^{d+1}L_a=N\,.
      \label{eq:intro-partition}
\end{equation}
We call this partition by $\rho$. The corresponding 3D $\mathcal N=4$ theory is usually denoted by $T_\rho(SU(N))$ in the literature.
It is claimed in \cite{CremonesiEtAl2014} that the monopole formula for $T_\rho(SU(N))$ has the closed-form (post-cancelled) expression
\begin{equation}
 H[{T_\rho}(SU(N))](t;z)
 =\frac{\prod_{q=L_1+1}^{N}(1-t^q)}{D_{\bdry}(t;z)}\,,
 \label{eq:intro-main-reproduction}
\end{equation}
with the denominator {\small
\begin{equation}
\begin{aligned}
D_{\bdry}(t;z)
={}&\prod_{i=1}^{L_1}\prod_{j=2}^{c_i}
\left(1-t^{L_j-i+1}\right)
\\
&\times\prod_{i=1}^{L_1}\prod_{1\leq k<j\leq c_i}
\left[1-\left(z_k z_{k+1}\cdots z_{j-1}\right)
 t^{(L_j+L_k)/2-i+1}\right]
\\
&\times\prod_{i=1}^{L_1}\prod_{1\leq k<j\leq c_i}
\left[1-\left(z_k z_{k+1}\cdots z_{j-1}\right)^{-1}
 t^{(L_j+L_k)/2-i+1}\right].
\end{aligned}
\label{eq:intro-bdry-denominator}
\end{equation}}

\noindent We will explicitly show how to construct the monopole formula \eqref{eq:intro-main-reproduction} from the quiver Yangian: the denominator \eqref{eq:intro-bdry-denominator} corresponds to the boundary-adapted quiver Yangian generators, while the numerator $\prod_{q=L_1+1}^{N}(1-t^q)$ is explained by the Casimir relations on the $A$-type Slodowy slice $\mathcal{S}^{\mathfrak{gl}_N}_\rho$ whose coordinates are identified with boundary-adapted quiver Yangian generators (up to fugacity-preserving coordinate transformations).

\medskip

The paper is organized as follows.  Section~\ref{sec:review} reviews Coulomb 
branch algebras, the monopole formula, the Hall--Littlewood result for
$T_\rho(SU(N))$, and the correspondence between the Coulomb branch algebra and the quiver Yangian.  In
section~\ref{sec:boundary-generators} we 
construct the boundary-adapted quiver Yangian generators  and show they correspond one-to-one to factors of $D_{\rm bdry}(t;z)$.  Section~\ref{sec:slodowy-reproduction} derives the relations predicted by the monopole formula as $U(N)$ Casimirs on the Slodowy slice $\mathcal{S}^{\mathfrak{gl}_N}_\rho$ and reconstructs the
monopole formula from the quiver Yangian. Examples will also be included. We end in section~\ref{sec:discussion} with discussion and a list of future problems. Finally appendix~\ref{appsec:neutral-cancel} provides computational details for the neutral cancellation, which is used in section~\ref{sec:slodowy-reproduction}.

 \section{Review}
\label{sec:review}

In this section, we first
review the Coulomb branch algebras of 3D $\mathcal N=4$ theories and their quantization, then review the monopole formula and the closed-form Hall--Littlewood formula for the $T_\rho(SU(N))$ theory. Finally, we end with the map between the (truncated shifted) quiver Yangian and the Coulomb branch algebra, which can be obtained from the algebra action on the $\frac{1}{2}$-BPS vortices in 3D $\mathcal N=4$ theories.

\subsection{Coulomb branches and their quantized algebras}
\label{subsec:review-coulomb}

Choosing an $\mathcal N=2$ subalgebra out of the 3D $\mathcal N=4$
supersymmetry algebra fixes a complex structure on the
hyperk\"ahler Coulomb branch $\cM_C$ \cite{BullimoreDimofteGaiotto2017}.  The corresponding chiral ring is the
commutative Poisson algebra $\mathbb C[\cM_C]$ of holomorphic functions on the Coulomb branch, generated by gauge-invariant polynomials in (complex)
vector-multiplet scalars and $\frac{1}{2}$-BPS monopole operators (possibly dressed by scalars in the residual gauge group).  The $\Omega$-rotation with parameter $\epsilon$
then deforms the Poisson product on $\mathbb C[\cM_C]$ to a noncommutative product \cite{Nekrasov_2010,Yagi:2014toa}, and we denote the quantized algebra by
$\mathbb C_\epsilon[\cM_C]$.

The quantized Coulomb branch algebra $\mathbb C_\epsilon[\cM_C]$ has a natural and faithful representation, given by $\frac 1 2$-BPS vortices of the 3D $\mathcal N=4$ theory. The vortex Hilbert space $\mathcal H_{\text{vortex}}$ serves as a Verma module of $\mathbb C_\epsilon[\cM_C]$ \cite{BullimoreEtAl2018}. Its geometric origin lies in the moduli of
non-Abelian vortices and their brane realizations
\cite{HananyTong2003,EtoEtAl2006,EtoEtAlReview2006}. 
We have used this module $\mathcal H_{\text{vortex}}$ to deduce the map between the Coulomb branch  algebra and the quiver Yangian, by viewing $\mathcal H_{\text{vortex}}$ also as a representation $\mathcal R_{\text{vortex}}$ of the quiver Yangian \cite{ChenLi2026}. In this paper, we will continue using this module, which encodes the information of the truncation on the quiver Yangian, to select the boundary-adapted generator basis in section~\ref{sec:boundary-generators}.

Let us also briefly review the mathematical construction of the quantized Coulomb branch algebra for 3D $\mathcal{N}=4$ theories, developed in \cite{Nakajima2016, BFN2018, BFN2019}.
 For a gauge group $G$ and matter in representation $\mathbf N$, one forms a
space $\mathcal R_{G,\mathbf N}$ over the \textit{affine Grassmannian} of the complexified gauge group $G_{\mathbb C}$, given by
\begin{equation}
    \text{Gr}_{G_{\mathbb C}}:=G_{\mathbb C}(\mathcal K)/G_{\mathbb C}(\mathcal O)\,,\quad \text{where } \mathcal{K}=\mathbb C((z)) \text{ and } \mathcal{O}=\mathbb C[[z]]\,,
\end{equation}
and defines the
Coulomb branch algebra as the $G_{\mathbb C}(\mathcal O)$-equivariant Borel–Moore homology:
\begin{equation}
     \mathcal A(G,\mathbf N)
 =\left(H_*^{G_{\mathbb C}(\mathcal O)}
   (\mathcal R_{G,\mathbf N}),\ \star\right)\,,
\end{equation}
with the well-defined convolution product $\star$ \cite{BFN2018}. The quantization is realized by turning on the loop-rotation $\mathbb C^\times$ with parameter $\hbar$, and the 
quantized Coulomb branch algebra is defined as the $(G_{\mathbb C}(\mathcal O)\rtimes\mathbb C^\times)$-equivariant Borel-Moore homology:
\begin{equation}
 \mathcal A_\hbar(G,\mathbf N)
 =\left(H_*^{G_{\mathbb C}(\mathcal O) \rtimes\mathbb C^\times}
   (\mathcal R_{G,\mathbf N}), \ \star\right)\,.
 \label{eq:bfn-algebra}
\end{equation}
For the explicit construction of the algebra and the product $\star$, see \cite{Nakajima2016,BFN2018}; a derived-Satake
refinement appears in \cite{BFNDerived2019}.  For a quiver of $ADE$ type, the
Coulomb branch is related to a transversal slice in the affine Grassmannian and
its quantization to the corresponding truncated shifted Yangian \cite{BFN2019,KWWY2012}.
Generator theorems and extensions to quivers with symmetrizers (e.g. $BCFG$-type quivers) sharpen the
scope of this statement \cite{Weekes2019,NakajimaWeekes2021}, while the study of the Jordan
quiver gives a distinct comparison with cyclotomic rational Cherednik algebras
\cite{KoderaNakajima2018}.  We refer to \cite{Kamnitzer2022,braden2022,KamnitzerEtAl2018} for the discussion of surrounding
symplectic duality and category-$\mathcal O$ structures.

In this paper, we will use both the physical and the mathematical constructions of the Coulomb branch algebra.
We use the latter to identify operators and the former to justify
the affine-slice geometric target, see sections~\ref{sec:boundary-generators} and \ref{sec:slodowy-reproduction}.  Agreement of their Hilbert series \cite{Nakajima2016} is the key bridge that leads to the quiver Yangian reproduction of the monopole formula.

\subsection{The monopole formula}
\label{subsec:review-monopole}

Given the gauge group $G$, let ${G^\vee}$ be the GNO (or Langlands) dual of $G$ and $\Gamma_{G^\vee}/W_G$ the lattice of GNO
cocharacters modulo the Weyl group $W_G$.  For a 3D $\mathcal N=4$ quiver gauge theory of good or ugly type (in the sense of \cite{GaiottoWitten2009}), the refined
monopole formula \cite{CremonesiHananyZaffaroni2014} is defined by
\begin{equation}
 H(t,z)
 =\sum_{m\in\Gamma_{G^\vee}/W_G}
 z^{J(m)}t^{\Delta(m)}P_G(t;m).
 \label{eq:monopole-general}
\end{equation}
The exponent $\Delta(m)$ is the $R$-charge of the bare monopole of
magnetic charge $m$ \cite{GaiottoWitten2009,Benna:2009xd,Bashkirov:2010kz}
:\footnote{\label{fn:u1r}This $R$-charge is defined with respect to the canonical $U(1)_R$ symmetry of 3D $\mathcal N=4$ SUSY. Given the standard R-symmetry $SU(2)_H\times SU(2)_C$, $U(1)_R$ is the diagonal Cartan combination, generated by $J^3_H+J^3_C$.}
\begin{equation}
    \Delta(m)={}-\sum_{\alpha\in\Phi_+(G)}
       \bigl|\langle\alpha,m\rangle\bigr| \\
  +\frac12\sum_I\sum_{w\in\mathcal R_I}
       \bigl|\langle w,m\rangle\bigr|,
\end{equation}
where $\alpha\in\Phi_+(G)$ is summed over the positive roots of $G$, while $w\in\mathcal R_I$ is summed over the weights of the matter field representation $\mathcal R_I$.
$J(m)$ denotes the topological charge of the monopole w.r.t.\ the topological symmetry $Z(G^\vee)$, with $z$ the corresponding fugacity. 
Finally, $P_G(t;m)$
counts invariant scalar dressings under the residual gauge group $H_m\subset G$ which is preserved by the monopole charge $m$.
If $d_i(m)$ are the degrees of the Casimir invariants of $H_m$, then
\begin{equation}
 P_G(t;m)=\prod_{i=1}^{\operatorname{rk}G}\frac{1}{1-t^{d_i(m)}}.
 \label{eq:monopole-dressing-general}
\end{equation}
The formula \eqref{eq:monopole-general} is a Hilbert series counting gauge-invariant operators built from dressed monopoles.

\medskip

From now on we will focus on $A$-type 3D $\mathcal N=4$ quiver gauge theories with bi-fundamental matters, specified by the quiver in \eqref{eq:intro-quiver}. 
Let
$m^{(a)}_1\geq\cdots\geq m^{(a)}_{N_a}$ be the ordered integral magnetic charges
at node $a$, and let $z_a$ be the corresponding $U(1)_a$-topological fugacity, satisfying the overall $U(1)$ constraint $z_0^N\prod_{a=1}^d z_a^{N_a}=1$. Note that we have already quotiented by the Weyl group $S_{N_a}\subset U(N_a)$ by fixing the non-increasing order for entries of $m^{(a)}=(m^{(a)}_1,\dots,m^{(a)}_{N_a})$. 
In this case, \eqref{eq:monopole-general} specializes to
\begin{equation}
 H^{\text{linear}}(t;z_1,\ldots,z_d)
 =\sum_{\{m^{(a)}\}}
 \left(\prod_{a=1}^{d}z_a^{\sum_{i=1}^{N_a}m_i^{(a)}}\right)
 t^{\Delta(m)}
 \prod_{a=1}^{d}P_{U(N_a)}(t;m^{(a)}),
 \label{eq:monopole-linear}
\end{equation}
where the $R$-charge is
\begin{align}
 \Delta(m)={}&
 \left(\frac12\sum_{i=1}^{N_1}\sum_{\alpha=1}^{N}|m_i^{(1)}|
 +\frac12\sum_{a=1}^{d-1}\sum_{i=1}^{N_a}
       \sum_{j=1}^{N_{a+1}}|m_i^{(a)}-m_j^{(a+1)}|\right)
 \nonumber\\
 &-\left(\sum_{a=1}^{d}\sum_{1\leq i<j\leq N_a}
       |m_i^{(a)}-m_j^{(a)}|\right).
 \label{eq:monopole-dimension-linear}
\end{align}
The two terms in \eqref{eq:monopole-dimension-linear} are respectively the
hypermultiplet and the vector-multiplet contributions.  If the equal
entries of $m^{(a)}$ occur in blocks of multiplicities
$\lambda_{a,1},\ldots,\lambda_{a,s_a}$ with $\lambda_{a,1}+\dots+\lambda_{a,s_a}=N_a$, then substituting the Casimir degrees of
$\prod_\beta U(\lambda_{a,\beta})$ into
\eqref{eq:monopole-dressing-general} gives
\begin{equation}
 P_{U(N_a)}(t;m^{(a)})
 =\prod_{\beta=1}^{s_a}\prod_{r=1}^{\lambda_{a,\beta}}
   \frac{1}{1-t^r}.
 \label{eq:monopole-dressing-unitary}
\end{equation}

Given the monopole formula, taking plethystic logarithm then organizes the rational Hilbert series into candidate
generator, relation, and higher-syzygy sectors
\cite{BenvenutiEtAl2006}.  When it terminates as a finite polynomial of complete-intersection form, a positive
monomial records a generator fugacity and a negative monomial records a relation
fugacity.  For a generic quiver, the series need not be of
complete-intersection form, and its plethystic logarithm can contain an infinite
alternating sequence.  
However, for the good $A$-type theory (i.e.\ the $T_\rho(SU(N))$ theory), which is the focus of this paper, its Coulomb branch is claimed to be a complete intersection \cite{CremonesiEtAl2014,Kumar2026}, as we will check later by explicitly working out its monopole formula using the quiver Yangian.

\subsubsection{The Hall--Littlewood expression for \texorpdfstring{$T_\rho(SU(N))$}{T-rho(SU(N))}}
\label{subsec:review-hl}

The good $A$-type theories considered in this paper are labeled by the partition
$\rho=L$ in \eqref{eq:intro-partition}.  
Let $\rho^T=c$ be the conjugate of $\rho$, specified by heights of columns of the Young diagram of $\rho$:
\begin{equation}
 c_i=\#\{a\in\{1,\ldots,d+1\}:L_a\geq i\},
 \qquad i=1,\ldots,L_1\,,
 \label{eq:column-height}
\end{equation}
and introduce variables $\{x_a\}$ by
\begin{equation}
 x_1=z_0,
 \qquad x_{a+1}=z_a{x_a}\quad (a=1,\ldots,d)\,,
 \label{eq:x-z-map}
\end{equation}
and the overall $U(1)$ constraint becomes
$\prod_{a=1}^{d+1}x_a^{L_a}=1$.  
It is claimed in \cite{CremonesiEtAl2014} that the monopole formula of $T_\rho(SU(N))$, defined by \eqref{eq:monopole-linear}, can be re-expressed in terms of Hall-Littlewood polynomials as follows:
\begin{equation}
 H[T_\rho(SU(N))](t;\vec{x})
 =(1-t)^N K_\rho(\vec{x};t)\,
   \Psi^{U(N)}_{(0,\ldots,0)}(\vec{x}t^{\frac{1}{2}\vec{\omega}_\rho};t),
 \label{eq:chmz-hl}
\end{equation}
and we explain this expression piece by piece.
\begin{itemize}
\item The factor $K_\rho(\vec{x};t)$ is defined by:
 \begin{equation}
 K_\rho(\vec{x};t)
 =\prod_{i=1}^{L_1}\prod_{j,k=1}^{c_i}
 \left(1-x_jx_k^{-1}
 t^{(L_j+L_k)/2-i+1}\right)^{-1}\,.
 \label{eq:chmz-k}
\end{equation}

\item $\vec{\omega}_\rho$ is a shorthand notation:
\begin{equation}
    \vec{\omega}_\rho=(\omega_{\rho_1},\dots,\omega_{\rho_{d+1}})\,,
\end{equation}
where each $\omega_{\rho_i}$ is a length-$\rho_i$ tuple:
\begin{equation}
    \omega_{\rho_i}=(\rho_i-1,\rho_i-3,\dots,3-\rho_i,1-\rho_i)\,.
\end{equation}
The notation $t^{\frac{1}{2}\vec{\omega}_\rho}$ hence represents 
\begin{equation}
    t^{\frac{1}{2}\vec{\omega}_\rho}=(t^{\frac{1}{2}\omega_{\rho_1}},\dots,t^{\frac{1}{2}\omega_{\rho_{d+1}}})\,,\quad \text{with }\  t^{\frac{1}{2}\omega_{\rho_i}}=(t^{\frac{\rho_i-1}{2}},t^{\frac{\rho_i-3}{2}},\dots,t^{\frac{3-\rho_i}{2}},t^{\frac{1-\rho_i}{2}})\,.
\end{equation}
Finally, the notation $\vec{x}t^{\frac{1}{2}\vec{\omega}_\rho}$ represents the length-$N$ tuple:
\begin{equation}
    \vec{x}t^{\frac{1}{2}\vec{\omega}_\rho}=(x_1t^{\frac{1}{2}\omega_{\rho_1}},\dots,x_{d+1}t^{\frac{1}{2}\omega_{\rho_{d+1}}})\,.
\end{equation}
\item $\Psi^{U(N)}_{(n_1,\ldots,n_N)}(x_1,\dots,x_N;t)$ is the Hall--Littlewood polynomial associated with $U(N)$ with charge $\vec n=(n_1,\ldots,n_N)$ (satisfying $n_1\geq\dots\geq n_N$), defined by
\begin{equation}
    \Psi^{U(N)}_{(n_1,\ldots,n_N)}(x_1,\dots,x_N;t)=\sum_{\sigma\in S_N}x_{\sigma_1}^{n_1}\dots x_{\sigma_N}^{n_N}\prod_{1\leq i<j\leq N}\frac{1-tx_{\sigma_i}^{-1}x_{\sigma_j}}{1-x_{\sigma_i}^{-1}x_{\sigma_j}}\,.
\end{equation}
In the case of $\vec n=(0,\dots,0)$, it satisfies a simple identity \cite{CremonesiEtAl2014}:
\begin{equation}
 (1-t)^N\Psi^{U(N)}_{(0,\ldots,0)}(x_1,\dots,x_N;t)
 =\prod_{q=1}^{N}(1-t^q).
 \label{eq:chmz-zero-hl}
\end{equation}
\end{itemize}

Let us rewrite \eqref{eq:chmz-hl} in variables $z_i$ and compute its plethystic logarithm for later use. Using
\eqref{eq:x-z-map} to split the double product in \eqref{eq:chmz-k}, since
\begin{equation}
 x_jx_k^{-1}=
 \begin{cases}
 \displaystyle\prod_{r=k}^{j-1}z_r,&j>k,\\[4pt]
 1,&j=k,\\[4pt]
 \displaystyle\prod_{r=j}^{k-1}z_r^{-1},&j<k,
 \end{cases}
 \label{eq:x-ratio-z-string}
\end{equation}
the positive part of the plethystic logarithm separates into
\begin{align}
 A_{\rm diag}
 &=\sum_{i=1}^{L_1}\sum_{j=1}^{c_i}t^{L_j-i+1},
 \label{eq:a-diag}\\
 A_+
 &=\sum_{i=1}^{L_1}\sum_{1\leq k<j\leq c_i}
 \left(\prod_{r=k}^{j-1}z_r\right)
 t^{(L_j+L_k)/2-i+1},
 \label{eq:a-plus}\\
 A_-
 &=\sum_{i=1}^{L_1}\sum_{1\leq k<j\leq c_i}
 \left(\prod_{r=k}^{j-1}z_r^{-1}\right)
 t^{(L_j+L_k)/2-i+1},
 \label{eq:a-minus}
\end{align}
where $A_{\rm{diag}}$ counts neutral coordinates (i.e.\ Cartan generators of the algebra), while $A_\pm$ count positive/negative coordinates (i.e.\ rising/lowering generators of the algebra).

On the other hand, the negative part of the plethystic logarithm comes from
\eqref{eq:chmz-zero-hl}, and is given by
\begin{equation}
    -\sum_{q=1}^N t^q\,.
\end{equation}
Substituting the positive and the negative parts into \eqref{eq:chmz-hl} gives
\begin{equation}
 H[T_\rho(SU(N))](t;z)
 =\PE\!\left[
 A_{\rm diag}+A_++A_- -\sum_{q=1}^{N}t^q
 \right].
 \label{eq:chmz-plethystic}
\end{equation}
Note that the good-theory inequalities in \eqref{eq:intro-partition} ensure that every positive exponent of $t$ in
\eqref{eq:a-diag}--\eqref{eq:a-minus} is positive.  Cancelling one
neutral positive term against one negative term at each degree
$1,\dots,L_1$ leads to the post-cancelled result in
\eqref{eq:intro-main-reproduction}. After taking plethystic exponential, the surviving positive terms give
$D_{\bdry}(t;z)$ while the surviving negative terms give
$\prod_{q=L_1+1}^{N}(1-t^q)$.

\subsection{Map between Coulomb branch algebras and Quiver Yangians}
\label{subsec:review-qy}

Let us now give a lightning review of the quiver Yangian  and the correspondence between the Coulomb branch algebra of 3D $\mathcal N=4$ theory of quiver $\rm Q$ and the (truncated shifted) quiver Yangian of triple quiver $\hat Q$ with canonical potential $\hat W$, proposed in \cite{ChenLi2026}. We refer to \cite{LiYamazaki2020,Li2024} for a comprehensive discussion on the quiver Yangian and its representation theory.

For a quiver $Q=(Q_0,Q_1)$ (where $Q_0$ is the node set while $Q_1$ is the arrow set) with potential $W$, the (shifted) quiver Yangian QY$(Q,W)$ is generated node by node by
raising, lowering, and Cartan generators
\begin{equation}
 e^{(a)}(z)=\sum_{n\geq0}\frac{e_n^{(a)}}{z^{n+1}},
 \qquad
 f^{(a)}(z)=\sum_{n\geq0}\frac{f_n^{(a)}}{z^{n+1}},
 \qquad
 \psi^{(a)}(z)=\sum_{n\geq 0}\frac{\psi_n^{(a)}}{z^{n+1+s^{(a)}}}\,,\quad a\in Q_0\,.
 \label{eq:qy-mode-expansion}
\end{equation}
Here $s^{(a)}$ is called the shift of the quiver Yangian. For the linear quiver shown in \eqref{eq:intro-quiver}, $s^{(a)}=2N_{a}-N_{a+1}-N_{a-1}=L_{a+1}-L_a$.\footnote{The quiver Yangian with non-zero shifts should be called the ``shifted quiver Yangian”, but since
in this paper generically they are shifted, below we will drop the qualifier ``shifted”.}

To the quiver $Q$ one can associate a potential $W$, which is a sum of cyclic monomials of arrows: 
\begin{equation}\label{eq:QY_potential}
	W=\sum_{d=1}^{F}\# M_d(\{I\})\,,\quad I\in Q_1\,,
\end{equation}
where each monomial $M_d(\{I\})$ ($d=1,2,\dots,F$) is a closed loop of arrows:
\begin{equation}
	M_d(\{I\})=I_1I_2\dots I_n\,\text{ with }\ t(I_i)=s(I_{i+1})\,,  \quad 1\leq i\leq n \ \text{ and }\ I_{n+1}:=I_1\,, 
\end{equation}
where $s,t:\: Q_1\to Q_0$ map an arrow to its source and target, respectively.
To each arrow $I\in Q_1$ we can assign a weight $h_I$, and the potential \eqref{eq:QY_potential} enforces the \textit{loop constraints} on the weights $\{h_I\}$:
\begin{equation}\label{eq:loop_cons}
	\sum_{I\in M_d}h_I=0\,,\qquad d=1,2,\dots,F\,.
\end{equation}

\medskip 
The quadratic relations of QY$(Q,W)$ are controlled by \textit{bonding factors}:
\begin{equation}\label{eq:QY_bond}
	\varphi^{b\Leftarrow a}(z)=\frac{\prod_{I\in\{b\to a\}}(z+h_I)}{\prod_{I\in\{a\to b\}}(z-h_I)}\,, \quad a,b\in Q_0\,,
\end{equation}
where $h_I$ are equivariant weights of arrows $I\in Q_1$, constrained by the potential $W$. The quadratic relations of QY$(Q,W)$ are given by \cite{LiYamazaki2020,Li2024}
\begin{equation}\label{eq:QY_quadratic_relations}
	\begin{aligned}
		\psi^{(a)}(z)\psi^{(b)}(w)&=\psi^{(b)}(w)\psi^{(a)}(z)\,,\\
		\psi^{(a)}(z)e^{(b)}(w)&\simeq \varphi^{a\Leftarrow b}(z-w)e^{(b)}(w)\psi^{(a)}(z)\,,\\
		\psi^{(a)}(z)f^{(b)}(w)&\simeq \varphi^{a\Leftarrow b}(z-w)^{-1}f^{(b)}(w)\psi^{(a)}(z)\,,\\
		e^{(a)}(z)e^{(b)}(w)&\sim (-1)^{|a||b|}\varphi^{a\Leftarrow b}(z-w)e^{(b)}(w)e^{(a)}(z)\,,\\
		f^{(a)}(z)f^{(b)}(w)&\sim (-1)^{|a||b|}\varphi^{a\Leftarrow b}(z-w)^{-1}f^{(b)}(w)f^{(a)}(z)\,,\\
		[e^{(a)}(z), f^{(b)}(w)\}&\sim -\delta_{ab}\frac{\psi^{(a)}(z)-\psi^{(b)}(w)}{z-w}\,,
	\end{aligned}
\end{equation}
\noindent where ``$\simeq$" means equality up to $z^nw^{m\geq0}$ terms, while ``$\sim$" means equality up to $z^{n\geq 0}w^{m}$ and $z^nw^{m\geq0}$ terms. $|a|$ is the $\mathbb{Z}_2$-grading of node $a$, and for quivers considered in this paper we can simply set $|a|=0$.

\medskip

It is conjectured in \cite{ChenLi2026} that the quantized Coulomb branch algebra of 3D $\mathcal N=4$ quiver gauge theory, specified by quiver $\rm Q$, can be captured by the (truncated shifted) quiver Yangian QY$^{\text{trun.}}(\hat Q,\hat W)$\footnote{The meaning of ``truncated" will be made manifest later.} of the triple quiver $\hat Q$ (with canonical potential $\hat W$), and this has been explicitly checked for tree-type quivers. Here a triple quiver $\hat Q=(\hat Q_0,\hat Q_1)$ is obtained from the original quiver $\rm Q=(\rm Q_0,\rm Q_1)$ by adding reverse arrows and self-loops for all nodes:
\begin{equation}
    \hat{Q}_0=\mathrm{Q}_0,\qquad \hat{Q}_1= \rm Q_1\cup \rm Q_1^{\text{reversed}}\cup \{\text{self-loops}\}\,,
\end{equation}
where
\begin{equation}
    \mathrm{Q}_1^{\text{reversed}}=\{I^{b\to a}\ |\ I^{a\to b} \in \mathrm{Q}_1\},\quad \{\text{self-loops}\}=\{I^{a\to a}\ |\ a\in \rm Q_0\}\,.
\end{equation}
For example, for the $A$-type quiver gauge theory in \eqref{eq:intro-quiver}, the corresponding triple quiver $\hat Q$ is 
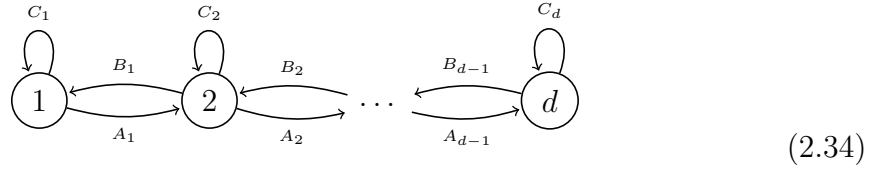
\begin{equation}
\begin{tikzpicture}
[->,auto=right, node distance=2cm,
			shorten >=1pt, semithick]
\node (v1) at (-2.75,0.2)[circle,draw] {1};
\node (v2) at (-0.5,0.2)[circle,draw] {2};
\node (e) at (1.75,0.15) {$\dots$};
\node (vd) at (4,0.2)[circle,draw] {$d$};

\draw (v2) edge [bend right=15] node {{\tiny $B_1$}} (v1);
\draw (v1) edge [bend right=15] node {{\tiny $A_1$}} (v2);
\draw (v1) edge [in=110,out=70,loop] node {{\tiny $C_1$}} (v1);
			
\draw (v2) edge [bend right=15] node {{\tiny $A_2$}} (e);
\draw (e) edge [bend right=15] node {{\tiny $B_2$}} (v2);
\draw (v2) edge [in=110,out=70,loop] node {{\tiny $C_2$}} (v2);

\draw (e) edge [bend right=15] node {{\tiny $A_{d-1}$}} (vd);
\draw (vd) edge [bend right=15] node {{\tiny $B_{d-1}$}} (e);
\draw (vd) edge [in=110,out=70,loop] node {{\tiny $C_{d}$}} (vd);

\end{tikzpicture}
\label{eq:A_qui_QY}
\end{equation}
and the associated canonical potential is given by
\begin{equation}\label{eq:A_triple_potential}
\widehat{W}=\sum_{a\geq1}^{d-1}\mathrm{Tr}\,(B_aC_aA_a-C_{a+1}B_aA_a)\,.
\end{equation}

In the case that $\rm Q$ is a tree-type quiver, after imposing the loop constraints of $\hat W$ there is  only one independent equivariant weight $\hbar$ for all arrows \cite{ChenLi2026}. For the $A$-type triple quiver in \eqref{eq:A_qui_QY}, the assignment of equivariant weights can be chosen as follows:
\begin{equation}
    h(A_a)=-\hbar\,,\quad h(B_a)=0\,,\quad h(C_a)=\hbar\,.
\end{equation}
Denote by $\mathfrak p(a)$ the parent node of node $a$ in the tree $\rm Q$, and denote by $\mathfrak{s}(a)$ the child of $a$.\footnote{The convention of orientation here is the same as that in \cite{ChenLi2026}: the tree grows in the direction away from the flavor node.} Then the bonding factors read
\begin{equation}\label{eq:tree_bond}
\varphi^{a\Leftarrow b}(z)=\left\{
\begin{array}{cc}
\frac{z+\hbar}{z-\hbar} \,,  &a=b\,,  \\
\frac{z}{z+\hbar} \,, & b=\mathfrak{p}(a)\,,\\
\frac{z-\hbar}{z}\,,& b= \mathfrak{s}(a)\,,\\
1\,,&a,b\text{ not adjacent}\,.
\end{array}\right.
\end{equation}
The correspondence between the quiver Yangian QY$(\hat{Q},\hat W)$ (specified by bonding factors \eqref{eq:tree_bond}) and the quantized Coulomb branch algebra $\mathbb C_\eps[\mathcal M_C]$ is given as follows.
\begin{itemize}
    \item First, identify  the quiver Yangian parameter with the $\Omega$-deformation parameter (or quantization parameter):
    \begin{equation}
        \hbar=\eps\,.
    \end{equation}
    \item The generators of $\mathbb C_\eps[\mathcal M_C]$ can be chosen as the following \cite{ChenLi2026, BullimoreEtAl2018}. For each $a\in\rm Q_0$ and $p\in \{1,\dots,N_a\}$, let $A^{(a)}_p$ denote the (first fundamental) minuscule cocharacter $(0,\dots, \stackunder{$1$}{$p$},\dots,0)$ of $U(N_a)$, and $\mathbb C_\eps[\mathcal M_C]$ has a triplet of generators:
    \begin{equation}
        \begin{aligned}
            \hat{v}^{(a)+}_p:&\text{ the (minuscule) monopole operator of charge } A^{(a)}_p 
            \,,\\
            \hat{v}^{(a)-}_p:&\text{ the (minuscule) monopole operator of charge } -A^{(a)}_p 
            \,,\\
            \hat{\varphi}^{(a)}_p:& \text{ the $p$-th component of the vector-multiplet complex scalar $\hat\varphi^{(a)}$}
            \,,
        \end{aligned}
        \label{eq:CBA_generators}
    \end{equation}
    which are rising, lowering, and Cartan generators, respectively.
    \item The map between $\mathbb C_\eps[\mathcal M_C]$ and QY$(\hat Q,\hat W)$ reads, for each $a\in \rm Q_0$:
    {\small
    \begin{align}
 e^{(a)}(z)
 &=\varepsilon^+(a)\sum_{p=1}^{N_a}
 \hat v_p^{(a)+}\frac{1}{z+\hat\varphi_p^{(a)}+\epsilon/2},
 \label{eq:chen-li-current-e}\\
 f^{(a)}(z)
 &=\varepsilon^-(a)\sum_{p=1}^{N_a}
 \hat v_p^{(a)-}\frac{1}{z+\hat\varphi_p^{(a)}+3\epsilon/2},
 \label{eq:chen-li-current-f}\\
 \psi^{(a)}(z)
 &=\frac{\displaystyle\prod_{b\in \mathfrak{s}(a)}\prod_{q=1}^{N_b}
 (z+\hat\varphi_q^{(b)}+\epsilon/2)
 \prod_{q=1}^{N_{\mathfrak{p}(a)}}
 (z+\hat\varphi_q^{(\mathfrak{p}(a))}+3\epsilon/2)}{\displaystyle\prod_{p=1}^{N_a}
 (z+\hat\varphi_p^{(a)}+\epsilon/2)
 (z+\hat\varphi_p^{(a)}+3\epsilon/2)},
 \label{eq:chen-li-current-psi}
\end{align}}
where the prefactors $\varepsilon^\pm(a)$ are defined by:
\begin{equation}\label{eq:epsilon+-}
\varepsilon^\pm(a):=\frac{e^{i\pi(\frac{N^{(a)}+N^{(\mathfrak{p}(a))}}{2}\mp\frac{1}{4})}}{\epsilon^{\frac{1}{2}}}\,.
\end{equation}
\end{itemize}

Expanding \eqref{eq:chen-li-current-e} and
\eqref{eq:chen-li-current-f} at large $z$ and comparing with
\eqref{eq:qy-mode-expansion} gives the explicit mode relations for $e$ and $f$ operators:
\begin{align}
 e_n^{(a)}
 &=\varepsilon^+(a)\sum_{p=1}^{N_a}
 \hat v_p^{(a)+}
 \left(-\hat\varphi_p^{(a)}-\frac{\epsilon}{2}\right)^n,
 \label{eq:chen-li-mode-e}\\
 f_n^{(a)}
 &=\varepsilon^-(a)\sum_{p=1}^{N_a}
 \hat v_p^{(a)-}
 \left(-\hat\varphi_p^{(a)}-\frac{3\epsilon}{2}\right)^n.
 \label{eq:chen-li-mode-f}
\end{align}
The Cartan mode relations are obtained by expanding
\eqref{eq:chen-li-current-psi} in a completely analogous manner, and each $\psi^{(a)}_n$ is a polynomial of order $n+1$ in $\hat{\varphi}$ operators. 
For example, the first 3 Cartan mode relations are given as follows. Define 
{\small
\begin{equation}
    \Lambda^{(a)}_r=\sum_{q=1}^{N_{\mathfrak p(a)}}\left(\hat\varphi^{\mathfrak p(a)}_q+\frac{3\eps}{2}\right)^r+\sum_{b\in \mathfrak s(a)}\sum_{q=1}^{N_b}\left(\hat\varphi^{(b)}_q+\frac{\eps}{2}\right)^r-\sum_{p=1}^{N_a}\left[\left(\hat\varphi^{(a)}_p+\frac{3\eps}{2}\right)^r+\left(\hat\varphi^{(a)}_p+\frac{\eps}{2}\right)^r\right]\,,
\end{equation} }
The map from $\psi^{(a)}_0,\psi^{(a)}_1,\psi^{(a)}_2$ to $\hat{\varphi}$'s reads:
\begin{equation}\label{eq:chen-li-mode-psi}
    \psi^{(a)}_0=\Lambda^{(a)}_1,\quad \psi^{(a)}_1=\frac{(\Lambda^{(a)}_1)^2-\Lambda^{(a)}_2}{2},\quad \psi^{(a)}_2=\frac{(\Lambda^{(a)}_1)^3-3\Lambda^{(a)}_1 \Lambda^{(a)}_2 +2 \Lambda^{(a)}_3}{6}\,.
\end{equation}
The higher-order Cartan mode relations, although more complicated, can be obtained by extracting the residues for higher-order poles in the large $z$ expansion of \eqref{eq:chen-li-current-psi}.

Note that relations 
\eqref{eq:chen-li-mode-e}--\eqref{eq:chen-li-mode-psi} are invariant w.r.t.\ the Weyl group, which is $S_{N_a}$ for each $U(N_a)$ factor. Hence, we see that the quiver Yangian generators are by their nature Weyl-invariant combinations of the dressed monopoles, and hence provide a natural basis for computing the monopole formula.

\medskip

In later sections, we will take the classical limit ($\hbar=\eps\to 0$) to simplify the calculations, see e.g.\ the calculation of Slodowy relations in section~\ref{sec:slodowy-reproduction}. 
Due to the appearance of $\epsilon^{-1/2}$ in the prefactors $\varepsilon^\pm(a)$ (see \eqref{eq:epsilon+-}), the map \eqref{eq:chen-li-mode-e}, \eqref{eq:chen-li-mode-f} will explode in this limit. Let us therefore define
the normalized $e,f$ modes (with $\hbar=\eps$):
\begin{equation}
 E_n^{(a)}=\hbar^{1/2}e_n^{(a)},
 \qquad
 F_n^{(a)}=\hbar^{1/2}f_n^{(a)}.
 \label{eq:normalized-modes}
\end{equation}
As we will see later, this rescaling is also essential when matching the quiver Yangian modes with the fugacities in
\eqref{eq:chmz-plethystic}.

\medskip

We conclude this section with the action of quiver Yangian generators on the $\frac 1 2$-BPS vortex Hilbert space $\mathcal H_{\text{vortex}}$ of the 3D $\mathcal N=4$ theory, which provides a faithful representation for the Coulomb branch algebra, see section~\ref{subsec:review-coulomb}. 
Let us specialize to $A$ type only. For the theory $T_\rho(SU(N))$ specified by the quiver in \eqref{eq:intro-quiver}, the vortex Hilbert space $\mathcal H_{\text{vortex}}$ is spanned by vortex states \cite{BullimoreEtAl2018,ChenLi2026}:
\begin{equation}
    |k\rangle:=|\{\vec{k}^{(1)},\dots,\vec{k}^{(d)}\}\rangle\,,
\end{equation}
where each $\vec{k}^{(a)}$ for $a=1,\dots, d$ is an $N_a$-dimensional vector $\vec{k}^{(a)}=(k^{(a)}_1,\dots,k^{(a)}_{N_a})$ of non-negative integral entries. In addition, these entries are constrained by the boundary conditions:
\begin{equation}\label{eq:bdry-condition}
     k^{(a)}_p\leq k^{(a+1)}_p,\qquad \ a=1,\dots,d-1,\quad p=1,\dots,N_{a+1}.
\end{equation}
A vortex that grows out of this boundary should be truncated from $\mathcal H_{\text{vortex}}$.

On the vortex $|k\rangle$, the action of $\mathbb C_\eps[\mathcal M_C]$ generators reads:\footnote{This action is the massless limit of the action studied in \cite{BullimoreEtAl2018,ChenLi2026}. In this paper, unless otherwise specified we will adopt this limit to simplify the calculations.}
\begin{equation}
\begin{aligned}
\hat{v}_{p}^{(a)+} |k\rangle&=\frac{\prod_{q=1}^{N_{a-1}}(\hat{\varphi}^{(a)}_{p}-\hat{\varphi}^{(a-1)}_{q})}{\prod_{q\neq p}^{N_{a}}(\hat{\varphi}^{(a)}_{p}-\hat{\varphi}^{(a)}_{q})}|k+\delta^{(a)}_{p}\rangle\,,\\
\hat{v}_{p}^{(a)-} |k\rangle&=\frac{\prod_{q=1}^{N_{a+1}}(\hat{\varphi}^{(a+1)}_{q}-\hat{\varphi}^{(a)}_{p})}{\prod_{q\neq p}^{N_{a}}(\hat{\varphi}^{(a)}_{q}-\hat{\varphi}^{(a)}_{p})}|k-\delta^{(a)}_{p}\rangle\,,\\
\hat{\varphi}^{(a)}_p|k\rangle&=-(k^{(a)}_p\eps+\frac{\eps}{2})|k\rangle \,,
\end{aligned}
\end{equation}
where by $\pm \delta^{(a)}_{p}$ we mean adding/subtracting 1 to/from the $p$-th entry of $\vec k^{(a)}$. For details of the derivation of the action above see \cite[section 3]{ChenLi2026}. Using the map between quiver Yangian generators and Coulomb branch algebra generators and the identification $\hbar=\eps$, we have the following (normalized) quiver Yangian action: 
{\small
\begin{equation}
    \begin{aligned}
        E^{(a)}_n|k\rangle&=
        \hbar^{n+N_{a-1}-N_a+1}\sum_{p=1}^{N_a} 
        \frac{(k^{(a)}_p)^n\prod_{q=1}^{N_{a-1}}(k^{(a-1)}_q-k^{(a)}_p-1)}{\prod_{q\neq p}^{N_{a}}(k^{(a)}_q-k^{(a)}_p-1)}|k+\delta^{(a)}_{p}\rangle\,,\\
        F^{(a)}_n|k\rangle&=
        \sigma_a\hbar^{n+N_{a+1}-N_a+1}\sum_{p=1}^{N_a} 
        \frac{(k^{(a)}_p-1)^n\prod_{q=1}^{N_{a+1}}(k^{(a)}_p-k^{(a+1)}_q-1)}{\prod_{q\neq p}^{N_{a}}(k^{(a)}_p-k^{(a)}_q-1)}|k-\delta^{(a)}_{p}\rangle\,,\\
    \end{aligned}
    \label{eq:normailzed-EF- action}
\end{equation}}

\noindent where we have absorbed all phases into a single factor $\sigma_a=(-1)^{N_{a-1}-N_a+1}$. The action for $\psi$ modes can again be obtained order by order. For example, for the first 3 modes we have:
{\small
\begin{equation}
\label{eq:psi-action}
    \psi^{(a)}_0|k\rangle=\lambda^{(a)}_1|k\rangle,\ \ \psi^{(a)}_1|k\rangle=\frac{(\lambda^{(a)}_1)^2-\lambda^{(a)}_2}{2}|k\rangle, \ \ \psi^{(a)}_2|k\rangle=\frac{(\lambda^{(a)}_1)^3-3\lambda^{(a)}_1 \lambda^{(a)}_2 +2 \lambda^{(a)}_3}{6}|k\rangle \,,
\end{equation}}
with 
{\small
\begin{equation}
    \lambda^{(a)}_r=\hbar^r\left(\sum_{q=1}^{N_{a-1}}(-k^{(a-1)}_q+1)^r+\sum_{q=1}^{N_{a+1}}(-k^{(a+1)}_q)^r-\sum_{p=1}^{N_{a}}\left[(-k^{(a)}_p+1)^r+(-k^{(a)}_p)^r\right]\right)\,.
\end{equation}}

\medskip

The vortex Hilbert space $\mathcal H_{\text{vortex}}$ furnishes a faithful representation of the \textit{truncated} quiver Yangian QY$^{\text{trun.}}(\hat{Q},\hat W)$, which is isomorphic to the quantized Coulomb branch algebra $\mathbb C_\eps[\mathcal M_C]$ \cite{ChenLi2026}.
In the next section, we will explicitly work out the truncation on the quiver Yangian algebra from its action on the representation $\mathcal H_{\text{vortex}}$. This is reminiscent of how the quiver Yangian was first defined as a representation-motivated algebra, e.g.\
as the BPS algebra for non-compact toric Calabi-Yau threefolds, bootstrapped from its action on the BPS states \cite{LiYamazaki2020}.

 \section{Boundary-adapted quiver Yangian generators}
\label{sec:boundary-generators}

We now turn the Hall--Littlewood expression of the $T_\rho(SU(N))$ monopole formula (see \eqref{eq:chmz-hl} and \eqref{eq:chmz-plethystic})
into a concrete operator-counting problem. In this section, we will first decipher the positive terms in the plethystic logarithm of \eqref{eq:chmz-plethystic}, and show that they can be reproduced by the \textit{boundary-adapted} quiver Yangian generators, which naturally preserve the boundary condition \eqref{eq:bdry-condition} of the representation $\mathcal H_{\text{vortex}}$.

\subsection{Independent generators of the truncated quiver Yangian}

Let us first note that only a finite number of $e,f,\psi$ modes are algebraically independent in the truncated algebra QY$^{\text{trun.}}(\hat{Q},\hat W)$. 
Indeed, for $e$ and $f$ modes, this is not unexpected from the map \eqref{eq:chen-li-mode-e} and \eqref{eq:chen-li-mode-f}. We take the case of $e$ modes as an example. Define the $N_a\times N_a$ matrix
\begin{equation}
    \Sigma^{(a)}=\mathrm{diag} (-\hat{\varphi}^{(a)}_1-\frac{\eps}{2},\dots,-\hat{\varphi}^{(a)}_{N_a}-\frac{\eps}{2})\,.
\end{equation}
Applying Cayley–Hamilton theorem over $\mathbb C_\eps[\mathcal M_C]\simeq\text{QY}^{\text{trun.}}(\hat{Q},\hat W)$ to the matrix $\Sigma^{(a)}$, we have the following standard recurrence relation
\begin{equation}
 (\Sigma^{(a)})^{n+N_a}=
(\Sigma^{(a)})^{n+N_a-1}\sigma_1^{(a)}-
 (\Sigma^{(a)})^{n+N_a-2}\sigma_2^{(a)}+\cdots+
 (-1)^{N_a+1}(\Sigma^{(a)})^n\sigma_{N_a}^{(a)},\quad n\geq 0\,,
 \label{eq:matrix-recurrence}
\end{equation}
where $\sigma^{(a)}_r$ is the $r$-th elementary symmetric polynomial of $\{-\hat{\varphi}^{(a)}_1-\frac{\eps}{2},\dots,-\hat{\varphi}^{(a)}_{N_a}-\frac{\eps}{2}\}$. 
The relation \eqref{eq:matrix-recurrence} is nothing but the characteristic polynomial of the matrix $\Sigma^{(a)}$, and by repeatedly applying this relation we can reduce any $(\Sigma^{(a)})^n$ with power $n\geq N_a$ into a linear combination of lower powers $\{I,\Sigma^{(a)},\dots,(\Sigma^{(a)})^{N_a-1}\}$
\begin{equation}
    (\Sigma^{(a)})^n=C_{n,0} +\Sigma^{(a)} C_{n,1} +\dots+  (\Sigma^{(a)})^{N_a-1}C_{n,N_a-1}\,, \quad n\geq N_a\,,
\end{equation}
with coefficients $C_{n,i}$ given by certain combinations of $\sigma^{(a)}_r$, possibly depending on $n$. 
For the $e$ modes specified by  \eqref{eq:chen-li-mode-e}, this means that we have
\begin{equation}
    e^{(a)}_n=e^{(a)}_0 C_{n,0} +e^{(a)}_1C_{n,1} +\dots+ e^{(a)}_{N_a-1}C_{n,N_a-1}\,,\quad n\geq N_a\,,
\end{equation}
from which we conclude that, for each $a\in \rm Q_0$, only the first $N_a$ modes of $e^{(a)}_n$ (and hence $E^{(a)}_n$) are algebraically independent generators. For the $f$ modes, the same argument still holds and we only need to replace $\Sigma^{(a)}$ by 
\begin{equation}
    \tilde{\Sigma}^{(a)}=\text{diag}(-\hat{\varphi}^{(a)}_1-\frac{3\eps}{2},\dots,-\hat{\varphi}^{(a)}_{N_a}-\frac{3\eps}{2})\,,
\end{equation}
and replace $\sigma^{(a)}_r$ by the $r$-th elementary symmetric polynomial of the diagonal elements of $\tilde{\Sigma}^{(a)}$. 
We therefore conclude that the independent normalized generators for the rising and the lowering sector of 
QY$^{\text{trun.}}(\hat{Q},\hat W)$ are given by:
\begin{equation}
    \{E^{(a)}_0,E^{(a)}_1,\dots,E^{(a)}_{N_a-1}\}\cup \{F^{(a)}_0,F^{(a)}_1,\dots,F^{(a)}_{N_a-1}\},\quad a\in\rm Q_0\,.
\end{equation}

Finally, for the Cartan sector, recall that each $\psi^{(a)}_n$ ($a\in\rm Q_0$ and $n=0,1,\dots$) is a symmetric polynomial of order $n+1$ in $\hat{\varphi}^{(a)}_p$ ($a\in\rm Q_0$ and $p=1,\dots,N_a$), see our discussion on Weyl-invariance in section~\ref{subsec:review-qy}. A basic fact about symmetric polynomials is that for each node $a\in\rm Q_0$, there are $N_a$ independent symmetric polynomials for the $N_a$ variables $\{\hat{\varphi}^{(a)}_1,\dots,\hat{\varphi}^{(a)}_{N_a}\}$, and a complete basis can be chosen as the aforementioned elementary symmetric polynomials. They are of orders $1,\dots, N_a$, respectively. As a result, we conclude that the independent $\psi$ modes must be of the same amount, and without loss of generality we can choose the following basis for the Cartan sector
\begin{equation}
    \{\psi^{(a)}_0,\psi^{(a)}_1,\dots,\psi^{(a)}_{N_a-1}\},\quad a\in \rm Q_0\,,
\end{equation}
which replace the role of the elementary symmetric polynomials.

\medskip

To summarize, we have chosen the complete set of independent generators for the truncated quiver Yangian QY$^{\text{trun.}}(\hat{Q},\hat W)$ to be:
\begin{equation}
    \{E^{(a)}_0,E^{(a)}_1,\dots,E^{(a)}_{N_a-1}\}\cup \{F^{(a)}_0,F^{(a)}_1,\dots,F^{(a)}_{N_a-1}\}\cup \{\psi^{(a)}_0,\psi^{(a)}_1,\dots,\psi^{(a)}_{N_a-1}\},\quad a\in\rm Q_0\,.
\end{equation}
Specializing to the linear quiver \eqref{eq:intro-quiver}, which is the focus of this paper, the independent generators are therefore
\begin{equation}
 \cG_{\raw}=\bigcup_{a=1}^{d}
 \{E_n^{(a)},F_n^{(a)},\psi_n^{(a)}:0\leq n\leq N_a-1\}\,.
 \label{eq:raw-seed}
\end{equation}
Here by the subscript ``raw" we mean these generators only provide raw seeds to the algebra QY$^{\text{trun.}}(\hat{Q},\hat W)$, and we have not imposed the boundary condition \eqref{eq:bdry-condition} from the representation $\mathcal H_{\text{vortex}}$ on the algebra yet.

\bigskip

Given the generators, let us pin down their $R$-charge $\Delta$ and the topological charges $J$. In fact, these values are already fixed by the relations:
\begin{equation}
    \begin{aligned}
        E^{(a)}_n &\sim \sum_{p=1}^{N_a} \hat{v}^{(a)+}_p(-\hat{\varphi}^{(a)}_p-\frac{\eps}{2})^n\,,\\
        F^{(a)}_n &\sim \sum_{p=1}^{N_a} \hat{v}^{(a)-}_p(-\hat{\varphi}^{(a)}_p-\frac{3\eps}{2})^n\,,\\
        \psi^{(a)}_n&\sim \text{Weyl-invariant polynomial of order $n+1$ in $\hat{\varphi}$'s}\,,
    \end{aligned}
\end{equation}
where ``$\sim$" means equality up to dimensionless numerical factors.
And the fugacity datum for the monopole operators $\hat{v}^{(a)\pm}_p$ and the vector-multiplet complex scalar $\hat{\varphi}^{(a)}_p$, defined in \eqref{eq:CBA_generators}, are given as follows \cite{CremonesiHananyZaffaroni2014}.
\begin{itemize}
    \item The monopole operators $\hat{v}^{(a)\pm}_p$ have topological charge $\pm 1$ at node $a$, respectively, and their common $R$-charge is
    \begin{equation}
 \Delta_a:=\Delta(\pm A^{(a)}_p)=\frac{N_{a-1}+N_{a+1}}{2}-(N_a-1)
 =1+\frac{L_a-L_{a+1}}{2}\,,
 \label{eq:minuscule-dimension}
\end{equation}
where $\Delta(m)$ is defined by \eqref{eq:monopole-dimension-linear} and $A^{(a)}_p$ is the corresponding (first fundamental) minuscule cocharacter $(0,\dots, \stackunder{$1$}{$p$},\dots,0)$ of $U(N_a)$.

\item The vector-multiplet complex scalars $\hat{\varphi}^{(a)}_p$ are neutral under the topological symmetry, and have $R$-charge 1.\footnote{Recall that the vector-multiplet complex scalars have charge $(0,1)$ w.r.t.\ $U(1)_H\times U(1)_C$ \cite{BullimoreEtAl2016,ChenLi2026}, and hence have charge $1+0=1$ under $U(1)_R$ (see also footnote~\ref{fn:u1r}).} The $\Omega$-deformation parameter $\eps$, viewed as a background value of vector-multiplet complex scalars \cite{BullimoreEtAl2018}, has the same fugacity.
\end{itemize}
Consequently, the fugacities of the truncated quiver Yangian generators are fixed to be:
\begin{equation}
 E_n^{(a)}:z_a t^{\Delta_a+n},\qquad
 F_n^{(a)}:z_a^{-1}t^{\Delta_a+n},\qquad
 \psi_n^{(a)}:t^{n+1},
 \qquad 0\leq n\leq N_a-1.
 \label{eq:sec31-fugacities}
\end{equation}
Note that due to the rescaling \eqref{eq:normalized-modes}, the fugacities of $E$ and $F$ operators get shifted by a factor of $t^{\frac 1 2}$. As we will see, it is the shifted operators that carry the correct fugacities as predicted by the monopole formula \eqref{eq:chmz-plethystic}.

\subsection{Incorporating boundary conditions}
\label{sec:bdry-adapted op}
The independent generators listed in \eqref{eq:raw-seed} provide a complete generator basis of the truncated quiver Yangian. However, they are only raw seeds of the algebra, and we still need to incorporate the boundary conditions \eqref{eq:bdry-condition} of the representation as an intrinsic property of the algebra itself. To this end, we need to make a redefinition to the algebra generators, such that their action automatically preserves the conditions \eqref{eq:bdry-condition}. 

Given a vortex state lying right on the boundary $k^{(a)}_p=k^{(a+1)}_p$, where $a$ is a quiver node and $p$ an integer between 1 and $N_{a+1}$, a rising action at the node $a$ is forbidden, as it would produce a state with $k^{(a)}_p>k^{(a+1)}_p$ which violates the condition \eqref{eq:bdry-condition}, see the action \eqref{eq:normailzed-EF- action}. To perform such a rising action safely, one must combine it with a rising action at the node $a+1$ at the same time, and cancel all the states that break the boundary. The appropriate combination is hence of the form:
\begin{equation}
    [E^{(a)}_n,E^{(a+1)}_m]\,.
\end{equation}
It is straightforward to check that its action generates only the boundary-preserving states, while any boundary-breaking state\footnote{Say, a state with increased $k^{(a)}_p$ and $k^{(a+1)}_q$ for some $q\neq p$, which still has $k^{(a)}_p>k^{(a+1)}_p$ (since $k^{(a+1)}_p$ is unchanged) and hence breaks the boundary.} is cancelled by the commutator structure. 

\medskip

With this simple example in mind, we propose the \textit{boundary-adapted} generators of the truncated quiver Yangian QY$^{\text{trun.}}(\hat{Q},\hat W)$ of good $A$-type quiver \eqref{eq:intro-quiver} as follows.
\begin{itemize}
    \item For each quiver node $b=d,\dots,1$ and a preceding node $a=b,\dots,1$, there are $L_{b+1}=N_b-N_{b+1}$ chains of inequalities to be satisfied:
    \begin{equation}
        k^{(a)}_p  \leq\dots\leq  k^{(b-1)}_p   \leq k^{(b)}_p\,,\quad p=N_{b+1}+1,\dots,N_b\,,
        \label{eq:bdry-chain}
    \end{equation}
    and we define the boundary-adapted rising operators
    \begin{equation}
        E_{a:b,r}=\hbar^{a-b}
 \sum_{\substack{n_a+\cdots+n_b=r\\ n_a,\ldots,n_b\geq0}}
 [\cdots[[E_{n_a}^{(a)},E_{n_{a+1}}^{(a+1)}],\ldots],E_{n_b}^{(b)}]\, ,
 \label{eq:positive-boundary-op}
    \end{equation}
    where $1\leq a\leq b\leq d$ and $0\leq r\leq L_{b+1}-1$. The index $r$ is referred to as the \textit{level} of the boundary-adapted operators. The factor $\hbar^{a-b}$ is introduced to balance the fugacity, see later.

\item The boundary-adapted lowering operators can be defined as the dual of the rising operators, which guarantee the lowering action does not break the boundary \eqref{eq:bdry-chain}:
\begin{equation}
     F_{a:b,r}=\hbar^{a-b}
 \sum_{\substack{n_a+\cdots+n_b=r\\ n_a,\ldots,n_b\geq0}}
 [\cdots[[F_{n_b}^{(b)},F_{n_{b-1}}^{(b-1)}],\ldots],F_{n_a}^{(a)}]\,,
 \label{eq:negative-boundary-op}
\end{equation}
where again we have $1\leq a\leq b\leq d$ and $0\leq r\leq L_{b+1}-1$.

\item Finally, the Cartan sector is selected from the raw Cartan modes by requiring their levels match with those of the positive and the negative sectors:
\begin{equation}
 H_{a,r}:=\psi_r^{(a)},\qquad
 1\leq a\leq d,\qquad 0\leq r\leq L_{a+1}-1.
 \label{eq:boundary-cartan-op}
\end{equation}
Unlike the rising and lowering operators, $\psi_r^{(a)}$ acts diagonally and
does not require a commutator to preserve the vortex boundary.  The range of the level $r$ can also be understood
from the following considerations. As we will see in section~\ref{sec:slodowy-reproduction}, upon the identification of the Coulomb branch of $T_{\rho}(SU(N))$ with the type-$A$ Slodowy slice (modulo Casimir relations), the Cartan part of the Coulomb branch algebra is thus to be identified with the neutral coordinates on the slice. There are $L_{a+1}$ independent neutral generators for each $a=1,\dots,d$, with fugacities $t,t^2,\dots,t^{L_{a+1}}$ respectively. This fixes the range $0\leq r\leq L_{a+1}-1$ for the Cartan generators $\psi^{(a)}_r$.
\end{itemize}

\medskip

In summary, for the linear quiver in \eqref{eq:intro-quiver}, we obtain the following boundary-adapted generators of the corresponding truncated quiver Yangian QY$(\hat{Q},\hat W)$:
\begin{equation}\label{eq:bdry-adapted gen}
    \cG_{\text{bdry}}=\cG_{\text{positive}}\cup \cG_{\text{negative}}\cup\cG_{\text{neutral}}\,,
\end{equation}
with
\begin{equation}\label{eq:bdry-op-set}
\begin{aligned}
    \cG_{\text{positive}}&=\{E_{a:b,r}:1\leq a\leq b\leq d,\ 0\leq r \leq L_{b+1}-1\}\,,\\
    \cG_{\text{negative}}&=\{F_{a:b,r}:1\leq a\leq b\leq d,\ 0\leq r \leq L_{b+1}-1\}\,,\\
    \cG_{\text{neutral}}&=\{H_{a,r}:1\leq a\leq d,\ 0\leq r \leq L_{a+1}-1\}\,.\\
\end{aligned}
\end{equation}

Now we are ready to show that these generators precisely match with the denominator of the monopole formula of $T_\rho(SU(N))$, or equivalently, the positive terms $A_\pm$ and $A_\text{diag}$ in the plethystic logarithm, which are given by \eqref{eq:a-diag}--\eqref{eq:a-minus}. 

\paragraph{Rising \& lowering sectors.}
Let us start from the $E_{a:b,r}$ operators. Define two index sets:
\begin{equation}
 \begin{aligned}
  \mathcal I_E
   &:=\left\{(a,b,r):1\leq a\leq b\leq d,\quad
                 0\leq r\leq L_{b+1}-1\right\},\\
  \mathcal I_+
   &:=\left\{(i,k,j):1\leq i\leq L_1,\quad
                 1\leq k<j\leq c_i\right\}.
 \end{aligned}
 \label{eq:E-index-sets}
\end{equation}
The first set labels the operators $E_{a:b,r}$ in \eqref{eq:bdry-op-set}, while the second set labels the summands of $A_+$ in \eqref{eq:a-plus}. We will show there exists a bijection between the two sets. 
Consider the map
\begin{equation}
 \Phi:\mathcal I_E\longrightarrow\mathcal I_+,\quad
 (a,b,r)\longmapsto (i,k,j)=(L_{b+1}-r,a,b+1).
 \label{eq:E-index-map}
\end{equation}
It is well-defined.  Indeed, $0\leq r\leq L_{b+1}-1$ gives
$1\leq i\leq L_{b+1}\leq L_1$.  Since $a\leq b$ and the partition is non-increasing, we have
$L_a\geq L_{b+1}\geq i$; by the definition of $c_i$ (see \eqref{eq:column-height}), this means $1\leq (k=a)<(j=b+1)\leq c_i$. Hence, $\Phi$ is a well-defined map from $\mathcal{I}_E$ to $\mathcal I_+$.

Conversely, for $(i,k,j)\in\mathcal I_+$ define
\begin{equation}
 \Phi^{-1}(i,k,j)=(a,b,r)=(k,j-1,L_j-i).
 \label{eq:E-index-inverse}
\end{equation}
The condition $j\leq c_i$ is equivalent to $L_j\geq i$, so
$0\leq r\leq L_j-1=L_{b+1}-1$.  Since $c_i\leq d+1$, the condition $1\leq k<j\leq c_i$ leads to $1\leq (a=k)\leq(b= j-1)\leq d$. Hence $\Phi^{-1}$ is a well-defined map from $\mathcal I_+$ to $\mathcal I_E$. It is easy to see
\eqref{eq:E-index-map} and \eqref{eq:E-index-inverse} are inverse to one
another.  Thus every term in \eqref{eq:a-plus} corresponds to one,
and only one, boundary-adapted rising operator $E_{a:b,r}$, and vice versa.

It remains to check the fugacity. 
Since every summand in the commutator \eqref{eq:positive-boundary-op} has level
$n_a+\cdots+n_b=r$,  using fugacities of row generators in
\eqref{eq:sec31-fugacities}, the topological charge and $t$-degree of $E_{a:b,r}$ are
\begin{equation}
 \left(\prod_{c=a}^{b}z_c\right),\qquad
 \sum_{c=a}^{b}\Delta_c+r+(a-b),
 \label{eq:E-degree-before-simplification}
\end{equation}
respectively.  Here $(a-b)$ is the contribution of the factors $\hbar^{a-b}$ (since $\hbar=\epsilon$ has fugacity $t$).  Substituting
$\Delta_c=1+(L_c-L_{c+1})/2$ (see \eqref{eq:minuscule-dimension}), we have 
\begin{equation}
 \begin{aligned}
 \sum_{c=a}^{b}\Delta_c+r+(a-b)&=1+\frac{L_a-L_{b+1}}{2}+r\,,\\
 &= \frac{L_j+L_k}2-i+1\,, \quad (i=L_{b+1}-r,\ k=a,\ j=b+1)
 \end{aligned}
 \label{eq:E-degree-match}
\end{equation}
where we have used the index map \eqref{eq:E-index-map}.  Therefore the boundary-adapted generator $E_{a:b,r}$ has exactly the fugacity
\begin{equation}
 E_{a:b,r}:
 \left(\prod_{c=k}^{j-1}z_c\right)
 t^{(L_j+L_k)/2-i+1},
 \qquad (i,k,j)=\Phi(a,b,r),
 \label{eq:E-fugacity-match}
\end{equation}
which is the corresponding summand in \eqref{eq:a-plus}.  

For the lowering sectors, the reasoning is similar. The result is that every term in \eqref{eq:a-minus}
corresponds to one and only one $F_{a:b,r}$ operator, and vice versa. The fugacity of $F_{a:b,r}$ is:
\begin{equation}
 F_{a:b,r}:
 \left(\prod_{c=k}^{j-1}z_c^{-1}\right)
 t^{(L_j+L_k)/2-i+1},
 \qquad (i,k,j)=\Phi(a,b,r)\,.
 \label{eq:F-fugacity-match}
\end{equation}

\paragraph{Cartan sector.} For the Cartan operators listed in \eqref{eq:boundary-cartan-op}, we can use a similar trick to show that they reproduce the (post-cancelled)  Cartan plethystic logarithm \eqref{eq:a-diag}. The summands in $A_{\text{diag}}$ are initially indexed by $1\leq i \leq L_1$ and $1\leq j\leq c_i$.  Reindex them by introducing
$r=L_j-i$, so that $0\leq r\leq L_j-1$ (recall that $j\leq c_i$ is equivalent to $L_j\geq i$).  For a fixed $r$, the allowed values for $(i,j)$
are exactly $(L_j-r,j)$ with
$j=1,\dots,c_{r+1}$, and therefore
\begin{equation}
 A_{\rm diag}
 =\sum_{r=0}^{L_1-1}c_{r+1}\,t^{r+1}.
 \label{eq:cartan-diag-reindex}
\end{equation}
After the cancellation of $L_1$ of these terms with fugacities $t,t^2,\dots,t^{L_1}$ (see the discussion below \eqref{eq:chmz-plethystic}), the
post-cancelled neutral contribution is
\begin{equation}
 A_{\rm diag}^{\rm post-cancelled}
 =\sum_{r=0}^{L_1-1}(c_{r+1}-1)t^{r+1}=\sum_{r=0}^{L_1-1}\sum_{a=1}^{c_{r+1}-1}t^{r+1}\,,
 \label{eq:cartan-post-cancellation}
\end{equation}
where we have introduced a new dummy index $a$. 

The corresponding index sets for $H_{a,r}$ and $A_{\rm diag}^{\rm post-cancelled}$ are
\begin{equation}
 \begin{aligned}
  \mathcal I_H
   &:=\left\{(a,r):1\leq a\leq d,\quad
                 0\leq r\leq L_{a+1}-1\right\},\\
  \mathcal I_{\rm diag}
   &:=\left\{(r,a):0\leq r\leq L_1-1,\quad
                 1\leq a\leq c_{r+1}-1\right\}.
 \end{aligned}
 \label{eq:cartan-index-sets}
\end{equation}
They are in bijection under the map
\begin{equation}
 \Phi_H:\mathcal I_H\longrightarrow\mathcal I_{\rm diag},
 \qquad (a,r)\longmapsto(r,a).
 \label{eq:cartan-index-map}
\end{equation}
Indeed, the partition ordering gives
\begin{equation}
 0\leq r\leq L_{a+1}-1
 \Longleftrightarrow
 L_{a+1}\geq r+1
\Longleftrightarrow
 a+1\leq c_{r+1}
\Longleftrightarrow
 a\leq c_{r+1}-1.
 \label{eq:cartan-range-equivalence}
\end{equation}
Moreover, since $H_{a,r}=\psi^{(a)}_r$, their fugacities are (see \eqref{eq:sec31-fugacities})
\begin{equation}
 H_{a,r}:t^{r+1},
 \qquad (r,a)=\Phi_H(a,r),
 \label{eq:cartan-fugacity-match}
\end{equation}
which is precisely the corresponding summand in \eqref{eq:cartan-post-cancellation}. 

\bigskip

Therefore, we conclude that for the linear quiver \eqref{eq:intro-quiver} of $T_\rho(SU(N))$, the boundary-adapted generators of the truncated quiver Yangian QY$(\hat{Q},\hat W)$, given in \eqref{eq:bdry-adapted gen}, precisely reproduce the (post-cancelled) denominator of the monopole formula \eqref{eq:intro-main-reproduction}.

 \section{Casimir relations and construction of the monopole formula}
\label{sec:slodowy-reproduction}

The denominator match in section~\ref{sec:boundary-generators} leaves the numerator or equivalently, the
negative terms in the plethystic logarithm of \eqref{eq:chmz-plethystic}, to be interpreted. In this section, we construct the (classical) Casimir relations from the $A$-type Slodowy slice. This approach is motivated by the shifted-Yangian and affine-Grassmannian-slice results
of \cite{MV2002,MirkovicVybornov2002,KWWY2012,KamnitzerEtAl2018,BrundanKleshchev2004}, as we will briefly review in section~\ref{sec:CB as Ss}. We then show these relations precisely match the numerator of the (post-cancelled) monopole formula of $T_\rho(SU(N))$.
Examples will be included, where we also provide a direct comparison of this method with the canonical enumeration method, and show that the relations obtained in two different ways are equivalent.

For simplicity, we will consider the classical case ($\hbar=\eps= 0$) only, where the quadratic relations \eqref{eq:QY_quadratic_relations} become trivial and the quiver Yangian reduces to a commutative algebra.\footnote{This can be seen by realizing the bonding factor \eqref{eq:tree_bond} becomes 1 in the classical limit.} Since the quantization of the Coulomb branch is a flat deformation \cite{BFN2018}, this will not change the Hilbert series.

\subsection{Coulomb branches as Slodowy slices}
\label{sec:CB as Ss}
Let us start with an overview of mathematical arguments that lead to our approach of studying the Coulomb branch algebras of 3D $\mathcal N=4$ theories through Slodowy slices. We consider the $A$-type case below. Interested readers may refer to \cite{Cabrera_2019} 
for the discussion of surrounding
mirror symmetry and generalizations to $BCD$ types.

\medskip

First, the mathematical construction of Coulomb branches of 3D $\mathcal N=4$ theories in \cite{BFN2019} has identified the Coulomb branches of framed $ADE$-type quivers with transverse slices in the affine Grassmannian (reviewed in section~\ref{subsec:review-coulomb}), and also identified the associated quantized algebra with the corresponding truncated shifted Yangian. Applied to the theory $T_\rho(SU(N))$ specified by the quiver in \eqref{eq:intro-quiver},
this gives the affine-Grassmannian-slice realization of the (classical) Coulomb branch \cite[Theorem~3.10]{BFN2019}:
\begin{equation}
    \mathcal M_C[T_\rho(SU(N))] \longleftrightarrow W^\lambda_\mu\,,
\end{equation}
where the affine-Grassmannian-slice $W^\lambda_\mu$ is specified by two coweights:
\begin{equation}
    \lambda=(N,0,\dots,0)\,,\quad\mu=\rho\,.
\end{equation}
The slice $W^\lambda_\mu$ has a graded Hilbert series given by the monopole formula \eqref{eq:chmz-hl}.

Second, from \cite{maffei2000} we know that for $A$-type quivers, there exists an isomorphism between the slices of affine Grassmannians\footnote{In \cite{maffei2000} the author used $A$-type Nakajima quiver varieties, which are identified (under certain stability condition) with affine Grassmannian slices in later work \cite{MV2002}. } and the Slodowy slices intersected with the nilpotent cone. Specialized to the affine-Grassmannian-slice $W^\lambda_\mu$, the isomorphism gives
\begin{equation}\label{eq:nakajima-slodowy}
    W^\lambda_\mu \longleftrightarrow \mathcal S^{\mathfrak{gl}_N}_\rho \cap \mathcal N_{\mathfrak{gl}_N}\,,
\end{equation}
where $\mathcal S^{\mathfrak{gl}_N}_\rho$ is the standard Slodowy slice of Lie algebra $\mathfrak{gl}_N$ associated with partition $\rho$, while $\mathcal{N}_{\mathfrak{gl}_N}$ is the nilpotent cone of $\mathfrak{gl}_N$, to be defined in section~\ref{subsec:slodowy-coordinates}.
It is worth noting that this isomorphism can also be seen at the algebraic level: on one hand,  the Slodowy slices are naturally quantized by finite $W$-algebras \cite{Premet2002, wang2010}, while the quantization of slices in $A$-type affine Grassmannians is provided by the corresponding $A$-type shifted Yangians \cite{KWWY2012}. On the other hand, for the $A$ type, it is known that the finite $W$-algebras and the shifted Yangians are manifestly isomorphic to each other \cite{BrundanKleshchev2004,BrundanKleshchev2006}. (For generalization to e.g.\ $BCD$ type, see \cite{luetal2025}.)

\medskip

Hence, based on the chain of the mathematical results given above, we naturally propose that the Coulomb branch of $T_\rho(SU(N))$ can be formulated as the $A$-type Slodowy slice intersected with nilpotent cone.
The $N$ classical relations predicted by \eqref{eq:chmz-plethystic}, conjectured to be $U(N)$ Casimir invariants \cite{CremonesiEtAl2014}, come from the intersection with the nilpotent cone $\mathcal{N}_{\mathfrak{gl}_N}$.
As we will see later, these relations will be given by the $N$ Casimir invariants of the Slodowy slice coordinate matrix. 
We will calculate these Casimir invariants below, and show that they have the correct fugacities as predicted by the monopole formula.

\subsection{$A$-type Slodowy slice and Casimirs}
\label{subsec:slodowy-coordinates}
We now provide the construction of the Slodowy slice \cite{Slodowy1980}. Given the Lie algebra $\mathfrak{gl}_N$ and a partition $\rho=L$ of $N$, let $J=\bigoplus_{a=1}^{d+1}J_{L_a}$ be the nilpotent Jordan matrix of partition
$\rho$, and let $f=\bigoplus_a f_{L_a}$ be the opposite nilpotent matrix. Explicitly, the blocks $J_{m}$ and $f_{m}$ are given by
\begin{equation}
    (J_{m})_{ij}=\delta_{i+1,j}\,,\quad (f_{m})_{ij}=\delta_{i,j+1}\,, \qquad 1\leq i,j\leq m\,.
\end{equation}
The Slodowy
slice $\mathcal S^{\mathfrak{gl}_N}_\rho$ is defiend as the affine space
\begin{equation}
    \mathcal S^{\mathfrak{gl}_N}_\rho=J+\mathfrak{gl}_N^{f}\,, \quad \text{with } \mathfrak{gl}_N^{f}=\ker(\operatorname{ad}f)
 =\{X\in\mathfrak{gl}_N\mid[f,X]=0\}.
\end{equation}
This is the
standard transverse-slice construction (see e.g.\ \cite{BrundanKleshchev2004}). The coordinates on the Slodowy slice $\mathcal S^{\mathfrak{gl}_N}_\rho$ can be organized into the coordinate matrix
\begin{equation}
 M=J+X,\qquad [X,f]=0\,,
 \label{eq:slodowy-matrix}
\end{equation}
where $J=\bigoplus_{a=1}^{d+1}J_{L_a}$ and $X$ is the centralizer of $f=\bigoplus_a f_{L_a}$. The matrix $X$ can be explicitly given as follows.
For block sizes $m$ and $n$, set $\ell=\min(m,n)$ and
$s=\max(m-n,0)$.  The equation $Xf=fX$ can then be solved, block by block, by the lower-Toeplitz matrices
$T_{m,n}(y_0,\ldots,y_{\ell-1})$:
\begin{equation}
 \bigl(T_{m,n}(y_0,\ldots,y_{\ell-1})\bigr)_{ij}
 =
 \begin{cases}
 y_{i-j-s},&s\leq i-j\leq s+\ell-1,\\
 0,&\text{otherwise},
 \end{cases}
 \qquad
 \substack{1\leq i\leq m,\ 1\leq j\leq n}.
 \label{eq:lower-toeplitz-block}
\end{equation}
For example
$T_{2,2}(y_0,y_1)=\left(\begin{smallmatrix}y_0&0\\y_1&y_0\end{smallmatrix}\right)$.

To display all coordinates at once, write
\begin{equation}
 \begin{aligned}
 \mathbf Y_a&=(Y_{a,0},\ldots,Y_{a,L_a-1}),\\
 \mathbf E_{a:b}&=(\tilde{E}_{a:b,0},\ldots,\tilde{E}_{a:b,L_{b+1}-1}),\\
 \mathbf F_{a:b}&=(\tilde{F}_{a:b,0},\ldots,\tilde{F}_{a:b,L_{b+1}-1}).
 \end{aligned}
 \label{eq:slodowy-coordinate-vectors}
\end{equation}
Then the full centralizer matrix $X$ is
\begin{equation}
 X=\begin{pmatrix}
 T_{L_1,L_1}(\mathbf Y_1)&T_{L_1,L_2}(\mathbf F_{1:1})&\cdots&T_{L_1,L_{d+1}}(\mathbf F_{1:d})\\
 T_{L_2,L_1}(\mathbf E_{1:1})&T_{L_2,L_2}(\mathbf Y_2)&\cdots&T_{L_2,L_{d+1}}(\mathbf F_{2:d})\\
 \vdots&\vdots&\ddots&\vdots\\
 T_{L_{d+1},L_1}(\mathbf E_{1:d})&T_{L_{d+1},L_2}(\mathbf E_{2:d})&\cdots&T_{L_{d+1},L_{d+1}}(\mathbf Y_{d+1})
 \end{pmatrix},
 \qquad M=\bigoplus_{a=1}^{d+1}J_{L_a}+X.
 \label{eq:slodowy-full-matrix}
\end{equation}
Equivalently, the blocks in \eqref{eq:slodowy-full-matrix} are
\begin{equation}
 X_{ba}=\begin{cases}
 T_{L_a,L_a}(\mathbf Y_a),&b=a,\\
 T_{L_b,L_a}(\mathbf E_{a:b-1}),&b>a,\\
 T_{L_b,L_a}(\mathbf F_{b:a-1}),&b<a.
 \end{cases}
 \label{eq:slodowy-block-coordinates}
\end{equation}

\medskip
Let us now characterize the $N$ relations predicted by the monopole formula \eqref{eq:chmz-plethystic} on the Slodowy slice. Recall that the Coulomb branch of $T_{\rho}(SU(N))$ is given by the intersection of the Slodowy slice $\mathcal S^{\mathfrak{gl}_N}_\rho$ and the nilpotent cone $\mathcal{N}_{\mathfrak{gl}_N}$, where the latter consists of all the nilpotent  elements of $\mathfrak{gl}_N$:
\begin{equation}
    \mathcal{N}_{\mathfrak{gl}_N}=\{A\in \mathfrak{gl}_N\mid A^k=0 \text{ for some }k\geq1 \}\,.
\end{equation}
Equivalently, the intersection with the nilpotent cone means the Slodowy slice matrix $M$ satisfies:
\begin{equation}
    \text{det}(\lambda I_N-M)=\lambda ^N\,.
\end{equation}
If we write 
\begin{equation}\label{eq:casimir-coeff}
    \text{det}(\lambda \mathbf 1_N-M)=\lambda ^N+C_1(M)\lambda^{N-1}+\dots+C_N(M)\,,
\end{equation}
where $C_i(M)$ is the $i$-th Casimir invariant of $M$, then the intersection with the nilpotent cone $ \mathcal{N}_{\mathfrak{gl}_N}$ leads to the following $N$ relations:
\begin{equation}
    C_1(M)=\dots=C_N(M)=0\,.
    \label{eq:casimir-relations}
\end{equation}
Below, we will show that these Casimir relations have  fugacities $t,\dots,t^N$, and hence match the prediction of the monopole formula \eqref{eq:chmz-plethystic}.

\subsection{Matching the monopole formula}
We now show that the monopole formula \eqref{eq:chmz-plethystic} can be precisely reproduced from the Slodowy slice matrix and the $N$ Casimir relations. 

Let us start by fixing a grading on the Slodowy slice coordinates \eqref{eq:slodowy-coordinate-vectors}, and show that the fugacities match.  
To this end, we choose the Kazhdan grading on $\mathcal S^{\mathfrak{gl}_N}_\rho$ \cite{gan2013,Ambrosio2023}, defined as follows. First, define the block-diagonal matrix
\begin{equation}
    h=\bigoplus_{a=1}^{d+1}h_{L_a},\quad \text{with }h_m=\text{diag}(m-1,m-3,\dots,3-m,1-m)\,.
\end{equation}
For each $a=1,\dots,d+1$, the blocks $J_{L_a},h_{L_a},f_{L_a}$ satisfy:
\begin{equation}
    [h_{L_a},J_{L_a}]=2J_{L_a},\quad [h_{L_a},f_{L_a}]=-2f_{L_a}\,,
\end{equation}
and hence we have 
\begin{equation}
    [h,J]=2J,\quad [h,f]=-2f\,.
\end{equation}
The Kazhdan action on the Slodowy slice matrix $M$ is defined by
\begin{equation}
 \kappa_u(M):=u^2\operatorname{Ad}_{u^{-h}}(M),
 \qquad M=J+X,\quad [X,f]=0.
 \label{eq:app-kazhdan-action}
\end{equation}
Because $\operatorname{Ad}_{u^{-h}}J=u^{-2}J$, the matrix $J$ is
preserved: $\kappa_u(J)=J$.  Moreover, since 
$\operatorname{Ad}_{u^{-h}}f=u^2f$, we see $[f,X]=0$ implies
$[f,\operatorname{Ad}_{u^{-h}}X]=0$.  Thus the action \eqref{eq:app-kazhdan-action}
preserves the affine slice $J+\mathfrak{gl}_N^{\,f}$.  

Now consider a lower-Toeplitz block
$T_{m,n}(y_0,\ldots,y_{\ell-1})$ from
\eqref{eq:lower-toeplitz-block}, with
$\ell=\min(m,n)$ and $s=\max(m-n,0)$.  The coordinate $y_r$ is called the
level-$r$ coordinate because it occupies exactly the matrix entries on the
diagonal
\begin{equation}
 i-j=s+r,\qquad r=0,\ldots,\ell-1.
 \label{eq:app-toeplitz-level}
\end{equation}
For example,
\begin{equation}
 T_{3,2}(y_0,y_1)=
 \begin{pmatrix}0&0\\y_0&0\\y_1&y_0\end{pmatrix};
 \qquad y_0\text{ has level }0,\quad y_1\text{ has level }1.
 \label{eq:app-toeplitz-example}
\end{equation}
Under the action of $h$, it is straightforward to see that the $(i,j)$ entry of $T_{m,n}(y_0,\ldots,y_{\ell-1})$ has the weight
\begin{equation}
 m-n-2(i-j).
 \label{eq:app-matrix-unit-weight}
\end{equation}
For a level-$r$ diagonal entry, this becomes (see \eqref{eq:app-toeplitz-level})
\begin{equation}
 m-n-2(s+r)=-|m-n|-2r.
 \label{eq:app-toeplitz-h-weight}
\end{equation}
Since the Kazhdan action uses $u^{-h}$, a level-$r$ entry is multiplied by
$u^{|m-n|+2r}$ under the adjoint factor. The action 
\eqref{eq:app-kazhdan-action} then gives
\begin{equation}
 \kappa_u(y_r)=u^{2+|m-n|+2r}y_r,
 \label{eq:app-toeplitz-fugacity}
\end{equation}
and if we set $t=u^2$, we have the $t$-degrees:
\begin{equation}
 \deg_t(y_r)=1+\frac{|m-n|}{2}+r.
 \label{eq:Kazhdan-t-degree}
\end{equation}

\medskip

\paragraph{Fugacities of slice coordinates.}
For the matrix $X$ in \eqref{eq:slodowy-full-matrix}, by \eqref{eq:Kazhdan-t-degree} we immediately find:
\begin{equation}
    \deg_t(\tilde{E}_{a:b,r})=\deg_t(\tilde{F}_{a:b,r})
 =1+\frac{L_a-L_{b+1}}{2}+r,
 \qquad r=0,\ldots,L_{b+1}-1\,,
 \label{eq:app-root-fugacity}
\end{equation}
which are precisely the $t$-degrees in \eqref{eq:E-fugacity-match} and \eqref{eq:F-fugacity-match} (see also \eqref{eq:E-degree-match}) for the charged coordinates. On the other hand, the neutral coordinates have $t$-degrees
\begin{equation}
 \deg_t(Y_{a,r})=r+1\,,
 \label{eq:app-neutral-fugacity}
\end{equation}
However, these $Y_{a,r}$ are not yet the post-cancelled neutral coordinates that we need. At each level $r=0,1,\dots,L_1-1$, there are $c_{r+1}$ candidates:
\begin{equation}
    Y_{1,r},\dots,Y_{c_{r+1},r}\,.
\end{equation}
However, after imposing the first $L_1$ Casimir relations, $C_1=\dots=C_{L_1}=0$, one neutral linear combination at each level is eliminated. (For a proof of this statement, see appendix~\ref{appsec:neutral-cancel}.) After this neutral cancellation, a natural basis for the remaining $c_{r+1}-1$ coordinates is given by the differences:
\begin{equation}
\label{eq:neutral-diff}
    \tilde{H}_{a,r}=Y_{a,r}-Y_{a+1,r}\,,\quad a=1,\dots,c_{r+1}-1\,.
\end{equation}
The coordinates $\tilde{H}_{a,r}$ are the post-cancelled neutral coordinates on the Slodowy slice, and they have $t$-degrees
\begin{equation}
 \deg_t(\tilde{H}_{a,r})=r+1\,.
 \label{eq:app-post-neutral-fugacity}
\end{equation}
If we fix the index $a$ of $\tilde{H}_{a,r}$, in order that the differences \eqref{eq:neutral-diff} are well defined, we necessarily have
\begin{equation}
        r\leq L_{a+1}-1\,,
\end{equation}
and hence the allowed range of level $r$ for $\tilde{H}_{a,r}$ (with $a$ fixed) is $0\leq r\leq L_{a+1}-1$. This explains the choice of index range in \eqref{eq:boundary-cartan-op}.

Finally, let us assign $z$-fugacities $z_az_{a+1}\dots z_b$ and $ (z_az_{a+1}\dots z_b)^{-1}$ to the charged coordinates $\tilde{E}_{a:b,r}$ and $\tilde{F}_{a:b,r}$, respectively. With these pinned down, the complete fugacities of the slice coordinates are

\begin{equation}\label{eq:slodowy-e-fugacity}
    \tilde{E}_{a:b,r}:\left(\prod_{c=a}^b z_c\right) t^{1+\frac{L_a-L_{b+1}}{2}+r}, \quad 1\leq a\leq b\leq d\,,\  0\leq r\leq L_{b+1}-1\,,
\end{equation}
\begin{equation}\label{eq:slodowy-f-fugacity}
    \tilde{F}_{a:b,r}:\left(\prod_{c=a}^b z_c\right)^{-1} t^{1+\frac{L_a-L_{b+1}}{2}+r}, \quad 1\leq a\leq b\leq d\,,\  0\leq r\leq L_{b+1}-1\,,
\end{equation}
and 
\begin{equation}\label{eq:slodowy-h-fugacity}
    \tilde{H}_{a,r}:t^{r+1}\,, \quad 0\leq r\leq L_{a+1}-1\,,
\end{equation}
which precisely match the (post-cancelled) positive terms of the plethystic logarithm in \eqref{eq:chmz-plethystic}.

\paragraph{Fugacities of Casimir relations.} It remains to show that the relation $C_i=0$ in \eqref{eq:casimir-relations} has fugacity $t^i$ for $i=1,\dots,N$. In fact, under the Kazhdan action \eqref{eq:app-kazhdan-action}, it is straightforward to see that
\begin{equation}
    \begin{aligned}
 \det\!\left(\lambda\mathbf 1_N-\kappa_u(M)\right)
 &=\det\!\left(\lambda\mathbf 1_N-u^2
       \operatorname{Ad}_{u^{-h}}M\right)\\
 &=u^{2N}\det\!\left(u^{-2}\lambda\mathbf 1_N-M\right)\\
 &=\lambda^N+\sum_{i=1}^{N}u^{2i}C_i(M)\lambda^{N-i}\,,
\end{aligned}
\label{eq:app-kazhdan-determinant-scaling}
\end{equation}
where we have used $\det\!\left(\operatorname{Ad}_{u^{-h}}M\right)=\det\!\left(u^{-h}M u^{h}\right)=\det\!\left(M\right)$. Comparing coefficients of $\lambda^{N-i}$ gives
\begin{equation}
 C_i(\kappa_u(M))=u^{2i}C_i(M)=t^iC_i(M)\,.
 \label{eq:app-kazhdan-Ci-degree}
\end{equation}
Hence, we see the Casimir relation $C_i(M)=0$ has the desired fugacity $t^i$, precisely matching the relation degree predicted by the monopole formula. After the neutral cancellation (to be discussed in appendix~\ref{appsec:neutral-cancel}), there are $N-L_1$ surviving Casimir relations $C_{L_1+1}=\dots=C_N=0$, which precisely match the (post-cancelled) negative terms in \eqref{eq:chmz-plethystic}.

\bigskip

We conclude that the formulation of the Coulomb branch of $T_\rho(SU(N))$ as the intersection of the Slodowy slice $\mathcal{S}^{\mathfrak{gl}_N}_\rho$  and the nilpotent cone $\mathcal{N}_{\mathfrak{gl}_N}$ leads to a complete interpretation of the (post-cancelled) monopole formula $H[T_\rho(SU(N))](t;z)$: the denominator of \eqref{eq:intro-main-reproduction} corresponds to the Slodowy slice coordinates \eqref{eq:slodowy-e-fugacity}--\eqref{eq:slodowy-h-fugacity}, while its numerator corresponds to $N-L_1$ Casimir relations $C_{L_1+1}=\dots=C_N=0$ defined in \eqref{eq:casimir-relations}.

\medskip

We make a final comment on the identification between the boundary-adapted quiver Yangian generators $\{E_{a:b,r},F_{a:b,r},H_{a,r}\}$ (given by \eqref{eq:positive-boundary-op}, \eqref{eq:negative-boundary-op}, \eqref{eq:boundary-cartan-op}) and the Slodowy slice coordinates $\{\tilde{E}_{a:b,r},\tilde{F}_{a:b,r},\tilde H_{a,r}\}$ (given by \eqref{eq:slodowy-coordinate-vectors}, \eqref{eq:neutral-diff}). Although they have identical index structures and fugacities, due to the freedom of coordinate choice, we cannot simply identify $E_{a:b,r}\leftrightarrow \tilde E_{a:b,r}$, $F_{a:b,r}\leftrightarrow \tilde F_{a:b,r}$, $H_{a,r}\leftrightarrow \tilde H_{a,r}$. In fact, $\{E_{a:b,r},F_{a:b,r},H_{a,r}\}$ and $\{\tilde{E}_{a:b,r},\tilde{F}_{a:b,r},\tilde H_{a,r}\}$ can be related by non-trivial coordinate transformations which, however, must be fugacity-preserving. This is to say, the boundary-adapted quiver Yangian generators $\{E_{a:b,r},F_{a:b,r},H_{a,r}\}$ and the Slodowy slice coordinates $\{\tilde{E}_{a:b,r},\tilde{F}_{a:b,r},\tilde H_{a,r}\}$ can be identified, only up to a fugacity-preserving coordinate transformation:
\begin{equation}
    \{E_{a:b,r},F_{a:b,r},H_{a,r}\} \ \xrightleftharpoons[\ \Psi^{-1}]{\Psi\ }\ \{\tilde{E}_{a:b,r},\tilde{F}_{a:b,r},\tilde H_{a,r}\}\,,\quad \text{where $\Psi$,$\Psi^{-1}$ preserve fugacities}\,.
\end{equation}
In the quiver Yangian language, the Casimir relations \eqref{eq:casimir-relations} can be written as 
\begin{equation}
    C_i(\Psi^{-1}(E_{a:b,r},F_{a:b,r},H_{a,r}))=0\,, 
\quad i=1,\dots,N\,,
\end{equation}
which, due to the fugacity-preserving property of $\Psi^{-1}$, still have fugacities $t,\dots,t^N$. This concludes the validity of constructing the monopole formula using the quiver Yangian.
We will explicitly see how this is done in examples below.

\subsection{Examples}
\subsubsection{$T_{(2,2)}(SU(4))$ of quiver $[U(4)]-(U(2))$}
Let us start from a simple example, the quiver $[U(4)]-(U(2))$, which has only one gauge node. The corresponding 3D $\mathcal{N}=4$ theory has $N=4$ and $L=c=(2,2)$. The post-cancelled monopole formula to be reproduced is 
\begin{equation}
 H_[T_{(2,2)}(SU(4))](t,z)=\frac{(1-t^3)(1-t^4)}{D_4(t,z)}\,,
 \label{eq:42-HS}
\end{equation}
with the denominator 
\begin{equation}
    D_4(t,z)=(1-t)(1-t^2)(1-zt)(1-z^{-1}t)(1-zt^2)(1-z^{-1}t^2)\,.
     \label{eq:42-denominator}
\end{equation}

Following the construction in section~\ref{sec:bdry-adapted op}, the boundary-adapted generators of the truncated quiver Yangian are simply the raw modes:
\begin{equation}
 \{E_0,E_1,F_0,F_1,H_0,H_1\}\,,
 \label{eq:42-modes}
\end{equation}
with fugacities:
\begin{equation}
    E_i: z t^{i+1}, \ \ \ F_i: z^{-1}t^{i+1}, \ \ \ H_i: t^{i+1},\quad i=0,1\,.
\end{equation}
Hence, we see these generators correspond one to one to the factors in the (post-cancelled) denominator \eqref{eq:42-denominator}, and the numerator of the \eqref{eq:42-HS} tells us there exist 2 relations among these generators, at orders $t^3$ and $t^4$, respectively. To find these relations, we can use two different methods.
\begin{itemize}
    \item \textbf{Canonical word enumeration.} In principle, given the truncated quiver Yangian QY$(\hat{Q},\hat W)$ and its faithful representation $\mathcal H_{\text{vortex}} $ (see section~\ref{subsec:review-qy}), one can always find the relations among generators by brute-force. 
    
    For example, to find the relation at order $t^3$, one can enumerate all the ``words" made by quiver Yangian generators at this order:
    \begin{equation}
       \{E_1F_0,E_0F_1,E_0H_0F_0,\hbar E_0F_0, H_1H_0,H_0^{\,3},\hbar H_1,\hbar H_0^{\,2},\hbar^2 H_0,\hbar^3\}\,,
    \end{equation}
    and search for the linear combination of them that acts as a zero operator on the representation $\mathcal H_{\text{vortex}}$. Given the action \eqref{eq:normailzed-EF- action} and \eqref{eq:psi-action}, this process can be easily automated using Mathematica. The resulting order-$t^3$ relation reads:
    \begin{align}
R_3^{\hbar\neq 0}=&E_1F_0+E_0F_1-\frac12E_0H_0F_0+\hbar E_0F_0
-\frac12 H_1H_0+\frac14H_0^3
\nonumber\\
&+\hbar H_1-\frac34\hbar H_0^2+\frac12\hbar^2H_0=0\,.
\label{eq:42-t3}
\end{align}
Similarly, the order-$t^4$ relation is found to be:
\begin{align}
R_4^{\hbar\neq 0}=&E_1F_1+\frac12E_0H_1F_0-\frac38E_0H_0^2F_0
+\frac34\hbar E_0H_0F_0-\frac12\hbar^2E_0F_0
\nonumber\\
&-\frac14 H_1^2+\frac5{64} H_0^4
+\frac54\hbar H_1H_0-\frac78\hbar H_0^3
-2\hbar^2H_1
\nonumber\\
&+\frac{33}{16}\hbar^2H_0^2-\frac32\hbar^3H_0+\frac14\hbar^4=0\,.
\label{eq:42-t4}
\end{align}
In the classical limit $\hbar\to0$, the two relations above smoothly reduce to 
\begin{align}
 R_3^{\rm QY}={}&E_1F_0+E_0F_1-\frac12E_0H_0F_0
 -\frac12H_1H_0+\frac14H_0^3=0\,,
 \label{eq:42-t3-classical}\\
 R_4^{\rm QY}={}&E_1F_1+\frac12E_0H_1F_0
 -\frac38E_0H_0^2F_0-\frac14H_1^2
 +\frac5{64}H_0^4=0\,.
 \label{eq:42-t4-classical}
\end{align}

\item \textbf{Slodowy slice matrix.} 
Let
\begin{equation}
 J_2=\begin{pmatrix}0&1\\0&0\end{pmatrix},\qquad
 B(x,y)=\begin{pmatrix}x&0\\y&x\end{pmatrix}.
 \label{eq:u24-blocks}
\end{equation}
The centralizer $X_4$ and the Slodowy slice matrix $M_4$  are given by
\begin{equation}
 X_4=\begin{pmatrix}
 B(Y_{1,0},Y_{1,1})&B(\tilde{F}_0,\tilde{F}_1)\\
 B(\tilde{E}_0,\tilde{E}_1)&B(Y_{2,0},Y_{2,1})
 \end{pmatrix},\qquad
 M_4=\operatorname{diag}(J_2,J_2)+X_4.
 \label{eq:u24-slice}
\end{equation}
Given $M_4$, the first 2 Casimir relations $C_1=C_2=0$ give the neutral cancellations
\begin{equation}
 Y_{1,0}+Y_{2,0}=0\,,\quad
 Y_{1,1}+Y_{2,1}=-2Y_{1,0}^2-2\tilde{E}_0\tilde{F}_0\,.
 \label{eq:u24-low-casimirs}
\end{equation}
We define the following basis for post-cancelled neutral coordinates
\begin{equation}
    \tilde{H}_{0}:=Y_{1,0}-Y_{2,0}\,,
 \quad \tilde{H}_{1}:=Y_{1,1}-Y_{2,1}\,,
\end{equation}
and the remaining Casimir relations are
\begin{align}
 C_3={}&-2\left(\tilde{E}_1\tilde{F}_0+\tilde{E}_0\tilde{F}_1\right)
 -\tilde{H}_{0}\tilde{H}_{1}=0,
 \label{eq:42-C3}\\
 C_4={}&\frac14\left(\tilde{H}_{0}^{\,2}+4\tilde{E}_0\tilde{F}_0\right)^2
 -\frac14\tilde{H}_{1}^{\,2}-\tilde{E}_1\tilde{F}_1=0.
 \label{eq:42-C4}
\end{align}
To compare the slice coordinates with the boundary-adapted quiver Yangian generators \eqref{eq:42-modes}, use the
fugacity-preserving coordinate change
\begin{align}
 \tilde{H}_{0}&=-\frac12H_0,
 &\tilde{H}_{1}&=-H_1+\frac12H_0^2+E_0F_0,
 \notag\\
 \tilde E_0&=E_0,
 &\tilde F_0&=-\frac14F_0,
 \notag\\
 \tilde E_1&=2E_1-E_0H_0,
 &\tilde F_1&=-\frac12F_1+\frac14H_0F_0.
 \label{eq:42-map}
\end{align}
Substituting \eqref{eq:42-map} into
\eqref{eq:42-C3} and \eqref{eq:42-C4} yields
\begin{equation}
 C_3=R_3^{\rm QY},\qquad
 C_4=R_4^{\rm QY}-\frac12H_0R_3^{\rm QY}\,,
 \label{eq:42-comparison}
\end{equation}
which are equivalent to relations obtained by brute-force word enumeration, \eqref{eq:42-t3-classical} and \eqref{eq:42-t4-classical}.

\end{itemize}

\subsubsection{$T_{(1,1,1)}(SU(3))$ of quiver $[U(3)]-(U(2))-(U(1))$}
Next, we consider the 3D $\mathcal N=4$ theory specified by quiver $[U(3)]-(U(2))_a-(U(1))_b$, where we denote the two gauge nodes of ranks 2 and 1 by $a$ and $b$, respectively. This theory has $N=3$, $L=(1,1,1)$, and $c=(3)$. The corresponding monopole formula is
\begin{equation}
 H[T_{(1,1,1)}(SU(3))](t,z_a,z_b)
 =\frac{(1-t^2)(1-t^3)}{D_3(t,z_a,z_b)}\,,
 \label{eq:321-HS}
\end{equation}
with
\begin{align}
 D_3(t,z_a,z_b)=&(1-t)^2(1-z_at)(1-z_a^{-1}t)
 (1-z_bt)(1-z_b^{-1}t)
 \notag\\[-2mm]
 &\times(1-z_az_bt)(1-z_a^{-1}z_b^{-1}t).
 \label{eq:321-denominator}
\end{align}
This theory is the simplest example with non-trivial boundary-adapted quiver Yangian generators. Define
\begin{equation}
\begin{aligned}
     &E_a=E^{(a)}_0\,,\quad F_a=F^{(a)}_0\,,\quad H_a=\psi^{(a)}_0\,,\\
     &E_b=E^{(b)}_0\,,\quad F_b=F^{(b)}_0\,,\quad H_b=\psi^{(b)}_0\,.\\
\end{aligned}
\end{equation}
By section~\ref{sec:bdry-adapted op}, there are now two boundary-adapted generators
\begin{equation}
    E_{a;b}=\hbar^{-1}[E_a,E_b]\,,\quad F_{a:b}=\hbar^{-1}[F_b,F_a]\,,
\end{equation}
and the complete set of boundary-adapted generators for the truncated quiver Yangian is
\begin{equation}
    \{E_a,E_b,E_{a:b},F_a,F_b,F_{a:b},H_a,H_b\}\,.
    \label{eq:321-bdry op}
\end{equation}
The fugacities of the generators are:
\begin{equation}
\begin{aligned}
     &E_a: z_at\,,\quad &&E_b: z_b t\,,\quad &&E_{a:b}:z_az_bt\,,\\
     &F_a: z_a^{-1}t\,,\quad &&F_b: z_b^{-1} t\,,\quad &&F_{a:b}:(z_az_b)^{-1}t\,,
\end{aligned}\label{eq:321-EF-fugacity}
\end{equation}
and
\begin{equation}\label{eq:321-H-fugacity}
    H_a:t\,,\quad H_b:t\,.
\end{equation}
With the fugacities \eqref{eq:321-EF-fugacity} and \eqref{eq:321-H-fugacity}, it is easy to see that the generators \eqref{eq:321-bdry op} can reproduce the monopole formula denominator \eqref{eq:321-denominator}. It remains to find the two relations at orders $t^2$ and $t^3$, which are predicted by the numerator of the monopole formula \eqref{eq:321-HS}.

\begin{itemize}
    \item \textbf{Canonical word enumeration.} By the brute-force enumeration method, we find precisely one order-$t^2$ relation and one order-$t^3$ relation:
    \begin{align}
R_2^{\hbar\neq 0}=&E_aF_a+E_bF_b-E_{a:b}F_{a:b}
-\frac13\left(H_a^2+H_aH_b+H_b^2\right)
\nonumber\\
&+\hbar(H_a+H_b)-\hbar^2=0\,,
\label{eq:321-t2}
\end{align}
and 
\begin{align}
R_3^{\hbar\neq 0}=&E_aE_bF_{a:b}+E_{a:b}F_aF_b
+E_a(2\hbar-H_b)F_a+E_b(H_a-2\hbar)F_b
\nonumber\\
&+\frac1{27}(H_b-H_a)^3=0\,.
\label{eq:321-t3}
\end{align}

In the classical limit $\hbar\to 0$, the two relations reduce to
\begin{align}
    R^{\rm QY}_2=&E_aF_a+E_bF_b-E_{a:b}F_{a:b}
-\frac13\left(H_a^2+H_aH_b+H_b^2\right)=0\,,\\
R_3^{\rm QY}=&E_aE_bF_{a:b}+E_{a:b}F_aF_b
-E_aH_bF_a+E_bH_aF_b+\frac1{27}(H_b-H_a)^3=0\,.
\end{align}

\item \textbf{Slodowy slice matrix.} For $L=(1,1,1)$ we have $J=0$, and the Slodowy slice matrix is given by
\begin{equation}
 M_3=X_3=\begin{pmatrix}
 \frac{2H_a+H_b}{3}&-F_a&F_{a:b}\\
 E_a&\frac{-H_a+H_b}{3}&-F_b\\
 E_{a:b}&E_b&\frac{-H_a-2H_b}{3}
 \end{pmatrix}\,,
 \label{eq:321-slice}
\end{equation}
where we have already imposed the neutral cancellation $C_1=0$ (which is equivalent to $\text{Tr}(M_3)=0$) and performed an appropriate fugacity-preserving coordinate transformation on the Slodowy slice.

The remaining Casimir invariants $C_2$ and $C_3$ are related to $R_2^{\rm QY}$ and $R_3^{\rm QY}$ by:
\begin{equation}
 C_2=R_2^{\rm QY},\qquad
 C_3=-R_3^{\rm QY}+\frac{H_a-H_b}{3}R_2^{\rm QY}.
 \label{eq:321-comparison}
\end{equation}

\end{itemize}

 \section{Discussion}
\label{sec:discussion}

In this work, we have reproduced the monopole formula \cite{CremonesiHananyZaffaroni2014,CremonesiEtAl2014} for good $A$-type 3D $\mathcal N=4$ quiver gauge theory, i.e.\ the theory $T_\rho(SU(N))$, from the truncated quiver Yangian, which is conjectured to be a reformulation of the 3D $\mathcal N=4$ Coulomb branch algebra \cite{ChenLi2026}. 
The calculation in sections~\ref{sec:boundary-generators}--\ref{sec:slodowy-reproduction} gives a direct interpretation of the two parts of the Hall--Littlewood expression of the monopole formula \cite{CremonesiEtAl2014}:  the positive monomials in its plethystic logarithm are the fugacities of boundary-adapted quiver Yangian generators, while the negative terms are the fugacities of the $U(N)$ Casimir invariants on the corresponding Slodowy slice, mapped back to the quiver Yangian via a fugacity-preserving coordinate transformation.  In this way, we have constructed the monopole formula of $T_\rho(SU(N))$ as the Hilbert series of the corresponding truncated quiver Yangian, and this result can be viewed as a necessary consistency check for the conjecture in \cite{ChenLi2026}.

The role of the boundary conditions \eqref{eq:bdry-condition} of the representation $\mathcal H_{\rm vortex}$,  originating from the physical vortex-module picture of \cite{BullimoreEtAl2018,HananyTong2003,EtoEtAl2006,EtoEtAlReview2006}, is essential. On one hand, as is shown in \cite{ChenLi2026}, they truncate the representation of the quiver Yangian QY$(\hat Q,\hat W)$ in such a way that the \textit{charge functions} have only simple poles, which is necessary for properly defining the quiver Yangian action and is not true for a general representation of QY$(\hat Q,\hat W)$. 
On the other hand, in this work the boundary conditions serve as the defining data of the truncation on the quiver Yangian: by translating the conditions \eqref{eq:bdry-condition} of the representation into an intrinsic property of the truncated quiver Yangian  QY$^{\text{trun.}}(\hat Q,\hat W)$, we obtain the boundary-adapted generators  whose fugacities completely reproduce those predicted by the monopole formula.

We have focused on $A$-type quivers in this work, for the following reasons. First, the Hall-Littlewood formula of the $T_\rho(SU(N))$ Hilbert series  has the  form of a complete intersection \cite{CremonesiEtAl2014} that makes the generator/relation split finite and clear.  For a general quiver, a comparably useful closed monopole formula is not known, and the Coulomb branch need not be a complete intersection. Second, type $A$ supplies particularly explicit descriptions \cite{MirkovicVybornov2002,MV2002,maffei2000,KWWY2012} of affine Grassmannian slices and Slodowy slices, which facilitate the study of 3D $\mathcal N=4$ Coulomb branches using shifted Yangians \cite{BullimoreDimofteGaiotto2017,BFN2019} and quiver Yangians \cite{ChenLi2026}. For the quiver Yangian, although its construction extends much further \cite{LiYamazaki2020,Bao2022,Bao2023,Li2024,GalakhovEtAl2024}, its Hilbert series cannot in general be organized by a finite numerator and denominator of the simple form used here.

\medskip

We end with some interesting problems for future research.
\begin{itemize}
    \item The first natural extension is to general tree-type quivers, where the linear construction of boundary-adapted generators (see e.g.\  \eqref{eq:bdry-chain} and \eqref{eq:positive-boundary-op}) must be replaced by a tree-dependent construction. Establishing the appropriate boundary-adapted generators would give a concrete consistency check of the present interpretation of the monopole formula beyond type $A$.
    \item The second extension is to $SO(N)$ and $Sp(N)$ gauge groups. The Hall-Littlewood expressions of monopole formulas exist for many classical groups \cite{CremonesiEtAl2014}, but the magnetic lattices and gauge invariants differ from the unitary case. Since the quiver Yangian provides an explicit generator/relation construction of the monopole formula, we are interested in how these discrepancies will manifest themselves in the quiver Yangian formalism.
    \item Finally, it would be useful to understand the construction within the broader BPS-algebra and cohomological Hall algebra framework \cite{KontsevichSoibelman2011}.  Such a formulation may provide a representation-independent origin for the boundary-adapted generators and the relations. 
\end{itemize}

We would like to report on these in future work.

\section*{Acknowledgments}
The author thanks Wei Li for helpful discussions.

\newpage
\appendix
\section{The first $L_1$ Casimirs account for neutral cancellation}
\label{appsec:neutral-cancel}
In this appendix, we will show that the first $L_1$ Casimir relations, $C_1=\dots=C_{L_1}=0$, lead to cancellations of the neutral coordinates $Y_{a,r}$. To be precise, for each level $r=0,\dots,L_1-1$, define the following combination:
\begin{equation}
 S_r=\sum_{a=1}^{c_{r+1}}(L_a-r)Y_{a,r}\,,\quad r=0,\dots,L_1-1\,.
 \label{eq:neutral-combination}
\end{equation}
We will show that the combination $S_r$ will be eliminated after imposing $C_{r+1}=0$.

Using Newton's identities, the Casimir invariants $C_i$ 
(see \eqref{eq:casimir-coeff}) and the traces $p_i=\text{Tr}(M^i)$ are related by:
\begin{equation}
 kC_k+\sum_{j=1}^{k}C_{k-j}p_j=0\,, \qquad k=1,\dots,N\,,\ \ \ C_0=1.
 \label{eq:newton-identities}
\end{equation}
After imposing $C_1=\cdots=C_{k-1}=0$, 
\eqref{eq:newton-identities} gives 
\begin{equation}
    C_k=-p_k/k\,,\qquad k=1,\dots,N\,. 
\end{equation}
We need the terms of $p_r$ ($r=1,\dots,L_1$) that are linear in the level-$(r-1)$ neutral coordinates $Y_{a,r-1}$ ($a=1,\dots,c_r$). Since $M=J+X$, such terms come from
\begin{equation}
    \sum_{l=0}^{r-1}\text{Tr}(J^l X J^{r-1-l})=r \text{Tr}(J^{r-1} X )\subset \text{Tr}(M^r)\,.
\end{equation}
Given the matrix $X$ (see \eqref{eq:slodowy-full-matrix}) and the nilpotent Jordan matrix $J$, in each block of size
$L_a$, the term in $p_r$ linear in $Y_{a,r-1}$ is
$r(L_a-r+1)Y_{a,r-1}$. Summing over all such blocks, we have
\begin{equation}
\begin{aligned}
    p_r&=r\sum_{a=1}^{c_r} (L_a-r+1)Y_{a,r-1}+\text{terms not linear in }\{Y_{a,r-1}\}\\
    &=rS_{r-1}+\text{terms not linear in }\{S_{r-1}\}\,,
\end{aligned}
\end{equation}
and hence
\begin{equation}
 C_r=-S_{r-1}+P_r(\tilde{E}_{a:b,r},\tilde{F}_{a:b,r},\tilde{H}_{a,r},S_0,\dots,S_{r-2}),
 \qquad r=1,\ldots,L_1,
 \label{eq:triangular-casimir}
\end{equation}
where $P_r$ depends only on the coordinates $\tilde{E}_{a:b,r},\tilde{F}_{a:b,r},\tilde{H}_{a,r}$ and on $S_r$ of lower levels.  The relations
\begin{equation}
 C_1=C_2=\cdots=C_{L_1}=0
 \label{eq:low-casimir-elimination}
\end{equation}
therefore eliminate $S_0,\ldots,S_{L_1-1}$ successively by setting
\begin{equation}
    S_0=P_1(\tilde{E}_{a:b,r},\tilde{F}_{a:b,r},\tilde{H}_{a,r})\,,\quad S_1=P_2(\tilde{E}_{a:b,r},\tilde{F}_{a:b,r},\tilde{H}_{a,r},S_0)\,,\dots \,,
\end{equation}
exactly one linear combination
at each level $r=0,\dots,L_1-1$.

\bibliography{references}
\bibliographystyle{utphys}

\end{document}